\documentclass[10pt]{article}
\usepackage{graphicx}
\usepackage{amsmath}
\usepackage{amssymb}
\usepackage{caption2}
\begin{document}
\begin{center}
{\large {\bf \sc{Strong decays of the vector tetraquark states with the masses about $4.5\,\rm{GeV}$  via the QCD sum rules  }}} \\[2mm]
Zhi-Gang  Wang \footnote{E-mail: zgwang@aliyun.com.  }     \\
 Department of Physics, North China Electric Power University, Baoding 071003, P. R. China
\end{center}

\begin{abstract}
We suppose that there exist three vector hidden-charm tetraquark states with the $J^{PC}=1^{--}$ at the energy about $4.5\,\rm{GeV}$, and investigate the two-body strong decays systematically. We obtain thirty QCD sum rules for the hadronic coupling constants based on rigorous quark-hadron duality, then obtain the partial decay widths, therefore the total widths approximately, which are  compatible with the experimental data of  the $Y(4500)$ from the BESIII collaboration. The $Y(4500)$ may be one vector tetraquark state having three main Fock components, or consists  of three different vector tetraquark states. We can search for the typical decays $ Y \to \frac{\bar{D}^0_1D^{0}-\bar{D}^{0}D^{0}_1}{\sqrt{2}}$,
 $\frac{\bar{D}^-_1D^{+}-\bar{D}^{-}D^{+}_1}{\sqrt{2}}$, $\frac{\bar{D}^0_0D^{*0}-\bar{D}^{*0}D^{0}_0}{\sqrt{2}}$,
$ \frac{\bar{D}^-_0D^{*+}-\bar{D}^{*-}D^{+}_0}{\sqrt{2}}$,
$\eta_c\omega$, $J/\psi\omega$ to diagnose the nature of the $Y(4500)$.
 \end{abstract}



\section{Introduction}
 In recent years, many charmonium-like states have been observed, especially the vector  charmonium-like states larger than $4.2\,\rm{GeV}$,  they  cannot be accommodated suitably   in the traditional quark model. In the present  work, we will focus on the $Y$ states and tetraquark picture. In 2022-2024, the BESIII collaboration observed three $Y$ structures at the energy about $4.5\,\rm{GeV}$ \cite{BESIII-Y4484-CPC,BESIII-Y4469-PRL,BESIII-Y4544-2401}.

In 2022, the BESIII collaboration measured the cross sections of the process $e^+e^- \to K^+K^-J/\psi$ at center-of-mass energies $\sqrt{s}=4.127-4.600\, \rm{GeV}$ based on the $15.6 \,\rm{fb}^{-1}$ data, and observed two resonant structures \cite{BESIII-Y4484-CPC}. The first structure has  a mass $4225.3\pm2.3\pm21.5\,\rm{MeV}$  and a width $72.9\pm6.1\pm30.8\,\rm{MeV}$, respectively, which are consistent with  the established $Y(4230)$. The second structure $Y(4500)$, which was observed for the first time with a statistical significance larger  than $8\sigma$,  has a mass $4484.7\pm13.3\pm24.1\,\rm{MeV}$  and a width $111.1\pm30.1\pm15.2\,\rm{MeV}$,  respectively.
In 2023, the BESIII collaboration measured the Born cross sections of the process $e^{+}e^{-}\to D^{*0}D^{*-}\pi^{+}$ at center-of-mass energies $\sqrt{s}=4.189-4.951 \,\rm{GeV}$, which corresponds to an integrated luminosity of $17.9\,{\rm fb}^{-1}$ data sample, and observed three enhancements \cite{BESIII-Y4469-PRL}. The resonances have the masses $4209.6\pm4.7\pm5.9\,{\rm MeV}$, $4469.1\pm26.2\pm3.6\,{\rm MeV}$ and $4675.3\pm29.5\pm3.5\,{\rm MeV}$ and widths  $81.6\pm17.8\pm9.0\,{\rm MeV}$, $246.3\pm36.7\pm9.4\,{\rm MeV}$ and $218.3\pm72.9\pm9.3\,{\rm MeV}$, respectively. They are consistent with the $Y(4230)$, $Y(4500)$ and $Y(4660)$ states, respectively.
In 2024, the BESIII collaboration measured the Born cross sections of the processes $e^+e^-\to\omega\chi_{c1}$ and $\omega\chi_{c2}$ at center-of-mass energies $\sqrt{s}=4.308-4.951\,\rm{GeV}$  with the data samples corresponding to an integrated luminosity of $11.0\,\rm{fb}^{-1}$ \cite{BESIII-Y4544-2401}. And they  observed the well-established resonance $\psi(4415)$ in the $\omega\chi_{c2}$ mass spectrum with  the mass $4413.6\pm9.0\pm0.8\, \rm{MeV}$ and width $110.5\pm15.0\pm2.9\, \rm{MeV}$, respectively. In addition, they  also observed a new resonance in the $\omega\chi_{c1}$ mass spectrum, and determined the mass and width to be $4544.2\pm18.7\pm1.7\, \rm{MeV}$ and $116.1\pm33.5\pm1.7\, \rm{MeV}$, respectively.

If we use the mass to represent the resonant structure, the $Y$ observed by the BESIII collaboration in 2022, 2023 and 2024 can be named as $Y(4484)$, $Y(4469)$ and $Y(4544)$, respectively,  they may be three different particles, or just one particle, the $Y(4500)$.

 Also in 2023, the BESIII collaboration studied the Born cross sections of the process $e^{+}e^{-}\to D_{s}^{\ast+}D_{s}^{\ast-}$  using the data samples at center-of-mass energies up to $4.95\, \rm{GeV}$, and observed two significant structures with the Breit-Wigner masses
 $4186.5\pm9.0\pm30\,\rm{MeV}$ and $4414.5\pm3.2\pm6.0\, \rm{MeV}$, respectively,  and widths  $55\pm17\pm53\,\rm{MeV}$ and $122.6\pm7.0\pm8.2\,\rm{MeV}$, respectively,  the third structure supports  a new state with  the Breit-Wigner mass $4793.3\pm 7.5\,\rm{MeV}$ and width $27.1\pm 7.0\,\rm{MeV}$, respectively  \cite{Y4790-BESIII}. Furthermore, the BESIII collaboration measured the cross section of the process $e^+e^-\to K^+K^-J/\psi$ using the data samples with an integrated luminosity of $5.85\, \rm{fb}^{-1}$ at center-of-mass energies $\sqrt{s}=4.61- 4.95\,\rm{GeV}$, and observed   a new resonance with a mass  $4708_{-15}^{+17}\pm21\,\rm{MeV}$ and a width  $126_{-23}^{+27}\pm30\,\rm{MeV}$ with a significance over $5\sigma$ \cite{Y4710-BESIII}.

There have been several explanations for the under-structures of the $Y(4500)$ et al.
In Refs.\cite{WangJZ-Y4500,WangJZ-Y4500-2},  the $Y(4500)$ is taken as the $5S-4D$ mixed charmonium state to study the three-body strong decays $Y(4500)\to J/\psi K^+K^-$ and $\pi^+D^0\bar{D}^{*-}$.
In an unquenched quark model including coupled-channel effects, the $Y(4500)$ and $Y(4710)$ are assigned to be the $3^3D_1$ and $4^3D_1$ charmonium states, respectively \cite{ZhongXH-Y4500}. In Ref.\cite{PengFZ-Y4500}, the $Y(4500)$ is assigned to be the $D_s\bar{D}_{s1}$ molecular state with the quantum numbers $J^{PC}=1^{--}$  based on the heavy-quark spin symmetry and light-flavor $SU(3)$ symmetry, such an assignment is also supported by calculating the masses  via the QCD sum rules \cite{SBTK-4500,WZG-Y4390-CPC} and Bethe-Salpeter equation \cite{FKGuo-DongXK}.

In the present work, we will focus on the picture  of tetraquark states.
In Refs.\cite{WZG-NPB-cucd-Vector,WZG-NPB-cscs-Vector},
we take the pseudoscalar, scalar, axialvector, vector, tensor (anti)diquarks as the elementary building blocks, and construct the vector and tensor four-quark currents without introducing explicit P-waves  to explore  the mass spectrum of the ground state vector hidden-charm tetraquark states in the framework of the QCD sum rules  comprehensively. At the energy about $4.5\,\rm{GeV}$, we obtain three hidden-charm tetraquark states with the quantum numbers  $J^{PC}=1^{--}$, the $[uc]_{\tilde{V}}[\overline{uc}]_{A}+[dc]_{\tilde{V}}[\overline{dc}]_{A}
-[uc]_{A}[\overline{uc}]_{\tilde{V}}-[dc]_{A}[\overline{dc}]_{\tilde{V}}$,
 $[uc]_{\tilde{A}}[\overline{uc}]_{V}+[dc]_{\tilde{A}}[\overline{dc}]_{V}
 +[uc]_{V}[\overline{uc}]_{\tilde{A}}+[dc]_{V}[\overline{dc}]_{\tilde{A}}$ and
 $[uc]_{S}[\overline{uc}]_{\tilde{V}}+[dc]_{S}[\overline{dc}]_{\tilde{V}}
 -[uc]_{\tilde{V}}[\overline{uc}]_{S}-[dc]_{\tilde{V}}[\overline{dc}]_{S}$ tetraquark states have
 the masses $4.53\pm0.07\, \rm{GeV}$, $4.48\pm0.08\,\rm{GeV}$ and $4.50\pm0.09\,\rm{GeV}$, respectively \cite{WZG-NPB-cucd-Vector}.
 They are all compatible with the $Y(4500)$ within uncertainties, there maybe exist three vector tetraquark states at the energy about $4.5\,\rm{GeV}$, or one vector tetraquark state which has three significant Fock components.
While the tetraquark states  $[sc]_{P}[\overline{sc}]_{A}-[sc]_{A}[\overline{sc}]_{P}$ and
$[sc]_{\tilde{V}}[\overline{sc}]_{A}-[sc]_{A}[\overline{sc}]_{\tilde{V}}$ with the quantum numbers $J^{PC}=1^{--}$ have the masses $4.80\pm0.08$ and  $4.70\pm0.08$, respectively, which support assigning them to be the  $Y(4790)$ and $Y(4710)$, respectively \cite{WZG-NPB-cscs-Vector}. For more works on the
spectroscopy, one can consult Refs. \cite{ChenZhu,WZG-HC-spectrum-PRD,WZG-tetra-psedo-NPB,ChenHX,Nielsen-review,WZG-HB-spectrum-EPJC,WZG-mole-IJMPA, ZhangJR-mole,QiaoCF-tetra-mole,Azizi-Review,WZG-tetra-cc-EPJC,WZG-XQ-mole-EPJA}.

In Ref.\cite{Y4500-decay-NPB}, we take the $Y(4500)$ as the  $[uc]_{\tilde{A}}[\overline{uc}]_{V}+[dc]_{\tilde{A}}[\overline{dc}]_{V}
  +[uc]_{V}[\overline{uc}]_{\tilde{A}}+[dc]_{V}[\overline{dc}]_{\tilde{A}}$ tetraquark state, and  explore the three-body decay $Y(4500)\to D^{*-}D^{*0}\pi^+$ in the framework of the light-cone QCD sum rules. It is the first time to use the light-cone QCD sum rules to calculate  the four-hadron coupling constants. And we observe that the process $Y(4500)\to D^{*-}D^{*0}\pi^+$  is not the main decay mode,  the partial decay width $6.43^{+0.80}_{-0.76}\,\rm{MeV}$ is too small to  match the experimental data.
The calculation is consistent with our naive expectation that the main decay channels are the two-body strong decays, and we  explore the two-body strong decays comprehensively in this work.

The article is arranged as follows:  we obtain  the QCD sum rules for the  hadronic coupling constants in section 2; in section 3, we present numerical results and discussions; section 4 is reserved for our conclusion.

\section{QCD sum rules for  the  hadronic coupling constants}
Firstly, we write down  the three-point correlation functions  in the  QCD sum rules,
\begin{eqnarray}
\Pi^{\bar{D}D\widetilde{A}V}_{\mu}(p,q)&=&i^2\int d^4xd^4y \, e^{ip\cdot x}e^{iq\cdot y}\, \langle 0|T\left\{J^{\bar{D}}(x)J^{D}(y)J_{-,\mu}^{\widetilde{A}V}{}^\dagger(0)\right\}|0\rangle\, ,
\end{eqnarray}

\begin{eqnarray}
\Pi^{\bar{D}^*D\widetilde{A}V}_{\alpha\mu}(p,q)&=&i^2\int d^4xd^4y \, e^{ip\cdot x}e^{iq\cdot y}\, \langle 0|T\left\{J_{\alpha}^{\bar{D}^*}(x)J^{D}(y)J_{-,\mu}^{\widetilde{A}V}{}^\dagger(0)\right\}|0\rangle\, ,
\end{eqnarray}

\begin{eqnarray}
\Pi^{\bar{D}^*D^*\widetilde{A}V}_{\alpha\beta\mu}(p,q)&=&i^2\int d^4xd^4y \, e^{ip\cdot x}e^{iq\cdot y}\, \langle 0|T\left\{J_{\alpha}^{\bar{D}^*}(x)J_\beta^{D^*}(y)J_{-,\mu}^{\widetilde{A}V}{}^\dagger(0)\right\}|0\rangle\, ,
\end{eqnarray}

\begin{eqnarray}
\Pi^{\bar{D}_0D^*\widetilde{A}V}_{\alpha\mu}(p,q)&=&i^2\int d^4xd^4y \, e^{ip\cdot x}e^{iq\cdot y}\, \langle 0|T\left\{J^{\bar{D}_0}(x)J_\alpha^{D^*}(y)J_{-,\mu}^{\widetilde{A}V}{}^\dagger(0)\right\}|0\rangle\, ,
\end{eqnarray}

\begin{eqnarray}
\Pi^{\bar{D}_1D\widetilde{A}V}_{\alpha\mu}(p,q)&=&i^2\int d^4xd^4y \, e^{ip\cdot x}e^{iq\cdot y}\, \langle 0|T\left\{J_{\alpha}^{\bar{D}_1}(x)J^{D}(y)J_{-,\mu}^{\widetilde{A}V}{}^\dagger(0)\right\}|0\rangle\, ,
\end{eqnarray}

\begin{eqnarray}
\Pi^{\eta_c\omega\widetilde{A}V}_{\alpha\mu}(p,q)&=&i^2\int d^4xd^4y \, e^{ip\cdot x}e^{iq\cdot y}\, \langle 0|T\left\{J^{\eta_c}(x)J_\alpha^{\omega}(y)J_{-,\mu}^{\widetilde{A}V}{}^\dagger(0)\right\}|0\rangle\, ,
\end{eqnarray}

\begin{eqnarray}
\Pi^{J/\psi\omega\widetilde{A}V}_{\alpha\beta\mu}(p,q)&=&i^2\int d^4xd^4y \, e^{ip\cdot x}e^{iq\cdot y}\, \langle 0|T\left\{J_{\alpha}^{J/\psi}(x)J_\beta^{\omega}(y)J_{-,\mu}^{\widetilde{A}V}{}^\dagger(0)\right\}|0\rangle\, ,
\end{eqnarray}

\begin{eqnarray}
\Pi^{\chi_{c0}\omega\widetilde{A}V}_{\alpha\mu}(p,q)&=&i^2\int d^4xd^4y \, e^{ip\cdot x}e^{iq\cdot y}\, \langle 0|T\left\{J^{\chi_{c0}}(x)J_\alpha^{\omega}(y)J_{-,\mu}^{\widetilde{A}V}{}^\dagger(0)\right\}|0\rangle\, ,
\end{eqnarray}

\begin{eqnarray}
\Pi^{\chi_{c1}\omega\widetilde{A}V}_{\alpha\beta\mu}(p,q)&=&i^2\int d^4xd^4y \, e^{ip\cdot x}e^{iq\cdot y}\, \langle 0|T\left\{J_\alpha^{\chi_{c1}}(x)J_\beta^{\omega}(y)J_{-,\mu}^{\widetilde{A}V}{}^\dagger(0)\right\}|0\rangle\, ,
\end{eqnarray}

\begin{eqnarray}
\Pi^{J/\psi f_0\widetilde{A}V}_{\alpha\mu}(p,q)&=&i^2\int d^4xd^4y \, e^{ip\cdot x}e^{iq\cdot y}\, \langle 0|T\left\{J_\alpha^{J/\psi}(x)J^{f_0}(y)J_{-,\mu}^{\widetilde{A}V}{}^\dagger(0)\right\}|0\rangle\, .
\end{eqnarray}
With the simple replacement $\widetilde{A}V \to \widetilde{V}A$, we obtain the
corresponding correlation functions for the current $J_{-,\mu}^{\widetilde{V}A}$.

\begin{eqnarray}
\Pi^{\bar{D}D S\widetilde{V}}_{\mu\nu}(p,q)&=&i^2\int d^4xd^4y \, e^{ip\cdot x}e^{iq\cdot y}\, \langle 0|T\left\{J^{\bar{D}}(x)J^{D}(y)J_{-,\mu\nu}^{S\widetilde{V}}{}^\dagger(0)\right\}
|0\rangle\, ,
\end{eqnarray}

\begin{eqnarray}
\Pi^{\bar{D}^*DS\widetilde{V}}_{\alpha\mu\nu}(p,q)&=&i^2\int d^4xd^4y \, e^{ip\cdot x}e^{iq\cdot y}\, \langle 0|T\left\{J_{\alpha}^{\bar{D}^*}(x)J^{D}(y)J_{-,\mu\nu}^{S\widetilde{V}}{}^\dagger(0)\right\}|0\rangle\, ,
\end{eqnarray}

\begin{eqnarray}
\Pi^{\bar{D}^*D^*S\widetilde{V}}_{\alpha\beta\mu\nu}(p,q)&=&i^2\int d^4xd^4y \, e^{ip\cdot x}e^{iq\cdot y}\, \langle 0|T\left\{J_{\alpha}^{\bar{D}^*}(x)J_\beta^{D^*}(y)J_{-,\mu\nu}^{S\widetilde{V}}{}^\dagger(0)\right\}|0\rangle\, ,
\end{eqnarray}

\begin{eqnarray}
\Pi^{\bar{D}_0D^*S\widetilde{V}}_{\alpha\mu\nu}(p,q)&=&i^2\int d^4xd^4y \, e^{ip\cdot x}e^{iq\cdot y}\, \langle 0|T\left\{J^{\bar{D}_0}(x)J_\alpha^{D^*}(y)J_{-,\mu\nu}^{S\widetilde{V}}{}^\dagger(0)\right\}|0\rangle\, ,
\end{eqnarray}

\begin{eqnarray}
\Pi^{\bar{D}_1DS\widetilde{V}}_{\alpha\mu\nu}(p,q)&=&i^2\int d^4xd^4y \, e^{ip\cdot x}e^{iq\cdot y}\, \langle 0|T\left\{J_{\alpha}^{\bar{D}_1}(x)J^{D}(y)J_{-,\mu\nu}^{S\widetilde{V}}{}^\dagger(0)\right\}|0\rangle\, ,
\end{eqnarray}

\begin{eqnarray}
\Pi^{\eta_c\omega S\widetilde{V}}_{\alpha\mu\nu}(p,q)&=&i^2\int d^4xd^4y \, e^{ip\cdot x}e^{iq\cdot y}\, \langle 0|T\left\{J^{\eta_c}(x)J_\alpha^{\omega}(y)J_{-,\mu\nu}^{S\widetilde{V}}{}^\dagger(0)\right\}|0\rangle\, ,
\end{eqnarray}

\begin{eqnarray}
\Pi^{J/\psi\omega S\widetilde{V}}_{\alpha\beta\mu\nu}(p,q)&=&i^2\int d^4xd^4y \, e^{ip\cdot x}e^{iq\cdot y}\, \langle 0|T\left\{J_{\alpha}^{J/\psi}(x)J_\beta^{\omega}(y)J_{-,\mu\nu}^{S\widetilde{V}}{}^\dagger(0)\right\}|0\rangle\, ,
\end{eqnarray}

\begin{eqnarray}
\Pi^{\chi_{c0}\omega S\widetilde{V}}_{\alpha\mu\nu}(p,q)&=&i^2\int d^4xd^4y \, e^{ip\cdot x}e^{iq\cdot y}\, \langle 0|T\left\{J^{\chi_{c0}}(x)J_\alpha^{\omega}(y)J_{-,\mu\nu}^{S\widetilde{V}}{}^\dagger(0)\right\}|0\rangle\, ,
\end{eqnarray}

\begin{eqnarray}
\Pi^{\chi_{c1}\omega S\widetilde{V}}_{\alpha\beta\mu\nu}(p,q)&=&i^2\int d^4xd^4y \, e^{ip\cdot x}e^{iq\cdot y}\, \langle 0|T\left\{J_\alpha^{\chi_{c1}}(x)J_\beta^{\omega}(y)J_{-,\mu\nu}^{S\widetilde{V}}{}^\dagger(0)\right\}|0\rangle\, ,
\end{eqnarray}

\begin{eqnarray}
\Pi^{J/\psi f_0S\widetilde{V}}_{\alpha\mu\nu}(p,q)&=&i^2\int d^4xd^4y \, e^{ip\cdot x}e^{iq\cdot y}\, \langle 0|T\left\{J_\alpha^{J/\psi}(x)J^{f_0}(y)J_{-,\mu\nu}^{S\widetilde{V}}{}^\dagger(0)\right\}|0\rangle\, ,
\end{eqnarray}
where the currents
\begin{eqnarray}
J^{\bar{D}}(x)&=&\bar{c}(x)i\gamma_{5} u(x)  \, ,\nonumber \\
J^{D}(y)&=&\bar{u}(y)i\gamma_{5} c(y) \, ,\nonumber \\
J_{\alpha}^{\bar{D}^*}(x)&=&\bar{c}(x)\gamma_{\alpha} u(x)  \, ,\nonumber \\
J_{\beta}^{D^*}(y)&=&\bar{u}(y)\gamma_{\beta} c(y) \, ,\nonumber \\
J^{\bar{D}_0}(x)&=&\bar{c}(x) u(x) \, ,\nonumber \\
J_{\alpha}^{\bar{D}_1}(x)&=&\bar{c}(x)\gamma_\alpha \gamma_5 u(x) \, ,
\end{eqnarray}
\begin{eqnarray}
J^{\eta_c}(x)&=&\bar{c}(x)i \gamma_5 c(x) \, ,\nonumber \\
J_{\alpha}^{J/\psi}(x)&=&\bar{c}(x)\gamma_{\alpha} c(x)  \, ,\nonumber \\
J^{\chi_{c0}}(x)&=&\bar{c}(x) c(x) \, ,\nonumber \\
J_{\alpha}^{\chi_{c1}}(x)&=&\bar{c}(x)\gamma_{\alpha}\gamma_5 c(x)  \, , \end{eqnarray}
\begin{eqnarray}
J_{\alpha}^{\omega}(y)&=&\frac{\bar{u}(y) \gamma_\alpha u(y)+\bar{d}(y) \gamma_\alpha d(y)}{\sqrt{2}} \, ,\nonumber \\
J^{f_0}(y)&=&\frac{\bar{u}(y)  u(y)+\bar{d}(y)  d(y)}{\sqrt{2}} \, ,
\end{eqnarray}
\begin{eqnarray}\label{AV-Current}
J_{-,\mu}^{\widetilde{A}V}(x)&=&\frac{\varepsilon^{ijk}\varepsilon^{imn}}{2}
\Big[u^{T}_j(x)C\sigma_{\mu\nu}\gamma_5 c_k(x)\bar{u}_m(x)\gamma_5\gamma^\nu C \bar{c}^{T}_n(x)+d^{T}_j(x)C\sigma_{\mu\nu}\gamma_5 c_k(x)\bar{d}_m(x)\gamma_5\gamma^\nu C \bar{c}^{T}_n(x) \nonumber \\
&&+u^{T}_j(x)C\gamma^\nu\gamma_5 c_k(x)\bar{u}_m(x)\gamma_5\sigma_{\mu\nu} C \bar{c}^{T}_n(x)+d^{T}_j(x)C\gamma^\nu\gamma_5 c_k(x)\bar{d}_m(x)\gamma_5\sigma_{\mu\nu} C \bar{c}^{T}_n(x) \Big] \, ,
\end{eqnarray}
\begin{eqnarray}\label{VA-Current}
J_{-,\mu}^{\widetilde{V}A}(x)&=&\frac{\varepsilon^{ijk}\varepsilon^{imn}}{2}
\Big[u^{T}_j(x)C\sigma_{\mu\nu} c_k(x)\bar{u}_m(x)\gamma^\nu C \bar{c}^{T}_n(x)
+d^{T}_j(x)C\sigma_{\mu\nu} c_k(x)\bar{d}_m(x)\gamma^\nu C \bar{c}^{T}_n(x)\nonumber\\
&&
-u^{T}_j(x)C\gamma^\nu c_k(x)\bar{u}_m(x)\sigma_{\mu\nu} C \bar{c}^{T}_n(x)
-d^{T}_j(x)C\gamma^\nu c_k(x)\bar{d}_m(x)\sigma_{\mu\nu} C \bar{c}^{T}_n(x) \Big] \, ,
\end{eqnarray}
\begin{eqnarray}\label{SV-Current}
J^{S\widetilde{V}}_{-,\mu\nu}(x)&=&\frac{\varepsilon^{ijk}\varepsilon^{imn}}{2}
\Big[u^{T}_j(x)C\gamma_5 c_k(x)  \bar{u}_m(x)\sigma_{\mu\nu} C \bar{c}^{T}_n(x)+d^{T}_j(x)C\gamma_5 c_k(x)  \bar{d}_m(x)\sigma_{\mu\nu} C \bar{c}^{T}_n(x)\nonumber\\
&&- u^{T}_j(x)C\sigma_{\mu\nu} c_k(x)  \bar{u}_m(x)\gamma_5 C \bar{c}^{T}_n(x) -d^{T}_j(x)C\sigma_{\mu\nu} c_k(x)  \bar{d}_m(x)\gamma_5 C \bar{c}^{T}_n(x)\Big] \, ,
\end{eqnarray}
 the $i$, $j$, $k$, $m$, $n$ are  color indexes \cite{WZG-NPB-cucd-Vector},   the $C$ is the charge conjugation matrix, the superscripts  $S$, $A$ and $V$ represent  the  scalar, axialvector and vector diquarks (or  antidiquarks), respectively, while  the $\widetilde{A}$ and $\widetilde{V}$ represent the axialvector and vector components in the tensor diquarks (or antidiquarks), respectively.

We insert  a complete set of intermediate hadronic states having non-vanishing  couplings  with the interpolating currents into the three-point correlation functions \cite{SVZ79, Reinders85}, and  isolate the ground state contributions explicitly,
\begin{eqnarray}
\Pi^{\bar{D}D\widetilde{A}V}_{\mu}(p,q)&=&\Pi_{\bar{D}D\widetilde{A}V}(p^{\prime2},p^2,q^2)
\,i\left(p-q\right)_\mu+\cdots\, ,
\end{eqnarray}

\begin{eqnarray}
\Pi^{\bar{D}^*D\widetilde{A}V}_{\alpha\mu}(p,q)&=&
\Pi_{\bar{D}^*D\widetilde{A}V}(p^{\prime2},p^2,q^2)
\,\left(-i\varepsilon_{\alpha\mu\lambda\tau}p^\lambda q^\tau\right)+\cdots\, ,
\end{eqnarray}

\begin{eqnarray}
\Pi^{\bar{D}^*D^*\widetilde{A}V}_{\alpha\beta\mu}(p,q)&=&
\Pi_{\bar{D}^*D^*\widetilde{A}V}(p^{\prime2},p^2,q^2)
\,\left(-ig_{\alpha\beta}p_\mu\right)+\cdots\, ,
\end{eqnarray}

\begin{eqnarray}
\Pi^{\bar{D}_0D^*\widetilde{A}V}_{\alpha\mu}(p,q)&=&
\Pi_{\bar{D}_0D^*\widetilde{A}V}(p^{\prime2},p^2,q^2)
\,\left(-ig_{\alpha\mu}p \cdot q\right)+\cdots\, ,
\end{eqnarray}

\begin{eqnarray}
\Pi^{\bar{D}_1D\widetilde{A}V}_{\alpha\mu}(p,q)&=&
\Pi_{\bar{D}_1D\widetilde{A}V}(p^{\prime2},p^2,q^2)
\,g_{\alpha\mu}+\cdots\, ,
\end{eqnarray}

\begin{eqnarray}
\Pi^{\eta_c\omega\widetilde{A}V}_{\alpha\mu}(p,q)&=&
\Pi_{\eta_c\omega\widetilde{A}V}(p^{\prime2},p^2,q^2)
\,\left(i\varepsilon_{\alpha\mu\lambda\tau}p^\lambda q^\tau\right)+\cdots\, ,
\end{eqnarray}

\begin{eqnarray}
\Pi^{J/\psi\omega\widetilde{A}V}_{\alpha\beta\mu}(p,q)&=&
\Pi_{J/\psi\omega\widetilde{A}V}(p^{\prime2},p^2,q^2)\,
\left(ig_{\alpha\beta}p_\mu\right)+\cdots\, ,
\end{eqnarray}

\begin{eqnarray}
\Pi^{\chi_{c0}\omega\widetilde{A}V}_{\alpha\mu}(p,q)&=&
\Pi_{\chi_{c0}\omega\widetilde{A}V}(p^{\prime2},p^2,q^2)
\,ig_{\alpha\mu}+\cdots\, ,
\end{eqnarray}

\begin{eqnarray}
\Pi^{\chi_{c1}\omega\widetilde{A}V}_{\alpha\beta\mu}(p,q)&=&
\Pi_{\chi_{c1}\omega\widetilde{A}V}(p^{\prime2},p^2,q^2)
\,\left(\varepsilon_{\alpha\beta\mu\lambda}p^\lambda \,p \cdot q\right)+\cdots\, ,
\end{eqnarray}

\begin{eqnarray}
\Pi^{J/\psi f_0\widetilde{A}V}_{\alpha\mu}(p,q)&=&
\Pi_{J/\psi f_0\widetilde{A}V}(p^{\prime2},p^2,q^2)
\,\left(-ig_{\alpha\mu}\right)+\cdots\, ,
\end{eqnarray}
other ground state contributions are given explicitly in the Appendix,
where
\begin{eqnarray}
\Pi_{\bar{D}D\widetilde{A}V}(p^{\prime2},p^2,q^2)&=&
\frac{\lambda_{\bar{D}D\widetilde{A}V}}{(m_Y^2-p^{\prime2})(m_D^2-p^2)(m_D^2-q^2)}
+\frac{C_{\bar{D}D\widetilde{A}V}}{(m_D^2-p^2)(m_D^2-q^2)}\nonumber\\
&&+\cdots\, ,
\end{eqnarray}
\begin{eqnarray}
\Pi_{\bar{D}^*D\widetilde{A}V}(p^{\prime2},p^2,q^2)&=&
\frac{\lambda_{\bar{D}^*D\widetilde{A}V}}{(m_Y^2-p^{\prime2})(m_{D^*}^2-p^2)(m_D^2-q^2)}
+\frac{C_{\bar{D}^*D\widetilde{A}V}}{(m_{D^*}^2-p^2)(m_D^2-q^2)}\nonumber\\
&&+\cdots\, ,
\end{eqnarray}
\begin{eqnarray}
\Pi_{\bar{D}^*D^*\widetilde{A}V}(p^{\prime2},p^2,q^2)&=&
\frac{\lambda_{\bar{D}^*D^*\widetilde{A}V}}{(m_Y^2-p^{\prime2})(m_{D^*}^2-p^2)(m_{D^*}^2-q^2)}
+\frac{C_{\bar{D}^*D^*\widetilde{A}V}}{(m_{D^*}^2-p^2)(m_{D^*}^2-q^2)}\nonumber\\
&&+\cdots\, ,
\end{eqnarray}
\begin{eqnarray}
\Pi_{\bar{D}_0D^*\widetilde{A}V}(p^{\prime2},p^2,q^2)&=&
\frac{\lambda_{\bar{D}_0D^*\widetilde{A}V}}{(m_Y^2-p^{\prime2})(m_{D_0}^2-p^2)(m_{D^*}^2-q^2)}
+\frac{C_{\bar{D}_0D^*\widetilde{A}V}}{(m_{D_0}^2-p^2)(m_{D^*}^2-q^2)}\nonumber\\
&&+\cdots\, ,
\end{eqnarray}
\begin{eqnarray}
\Pi_{\bar{D}_1D\widetilde{A}V}(p^{\prime2},p^2,q^2)&=&
\frac{\lambda_{\bar{D}_1D\widetilde{A}V}}{(m_Y^2-p^{\prime2})(m_{D_1}^2-p^2)(m_{D}^2-q^2)}
+\frac{C_{\bar{D}_1D\widetilde{A}V}}{(m_{D_1}^2-p^2)(m_{D}^2-q^2)}\nonumber\\
&&+\cdots\, ,
\end{eqnarray}

\begin{eqnarray}
\Pi_{\eta_c\omega \widetilde{A}V}(p^{\prime2},p^2,q^2)&=&
\frac{\lambda_{\eta_c\omega\widetilde{A}V}}{(m_Y^2-p^{\prime2})
(m_{\eta_c}^2-p^2)(m_{\omega}^2-q^2)}
+\frac{C_{\eta_c\omega\widetilde{A}V}}{(m_{\eta_c}^2-p^2)(m_{\omega}^2-q^2)}\nonumber\\
&&+\cdots\, ,
\end{eqnarray}

\begin{eqnarray}
\Pi_{J/\psi\omega \widetilde{A}V}(p^{\prime2},p^2,q^2)&=&
\frac{\lambda_{J/\psi\omega\widetilde{A}V}}{(m_Y^2-p^{\prime2})
(m_{J/\psi}^2-p^2)(m_{\omega}^2-q^2)}
+\frac{C_{J/\psi\omega\widetilde{A}V}}{(m_{J/\psi}^2-p^2)(m_{\omega}^2-q^2)}\nonumber\\
&&+\cdots\, ,
\end{eqnarray}
\begin{eqnarray}
\Pi_{\chi_{c0}\omega \widetilde{A}V}(p^{\prime2},p^2,q^2)&=&
\frac{\lambda_{\chi_{c0}\omega\widetilde{A}V}}{(m_Y^2-p^{\prime2})
(m_{\chi_{c0}}^2-p^2)(m_{\omega}^2-q^2)}
+\frac{C_{\chi_{c0}\omega\widetilde{A}V}}{(m_{\chi_{c0}}^2-p^2)(m_{\omega}^2-q^2)}\nonumber\\
&&+\cdots\, ,
\end{eqnarray}
\begin{eqnarray}
\Pi_{\chi_{c1}\omega \widetilde{A}V}(p^{\prime2},p^2,q^2)&=&
\frac{\lambda_{\chi_{c1}\omega\widetilde{A}V}}{(m_Y^2-p^{\prime2})
(m_{\chi_{c1}}^2-p^2)(m_{\omega}^2-q^2)}
+\frac{C_{\chi_{c1}\omega\widetilde{A}V}}{(m_{\chi_{c1}}^2-p^2)(m_{\omega}^2-q^2)}\nonumber\\
&&+\cdots\, ,
\end{eqnarray}
\begin{eqnarray}
\Pi_{J/\psi f_0 \widetilde{A}V}(p^{\prime2},p^2,q^2)&=&
\frac{\lambda_{J/\psi f_0\widetilde{A}V}}{(m_Y^2-p^{\prime2})
(m_{J/\psi}^2-p^2)(m_{f_0}^2-q^2)}
+\frac{C_{J/\psi f_0\widetilde{A}V}}{(m_{J/\psi}^2-p^2)(m_{f_0}^2-q^2)}\nonumber\\
&&+\cdots\, .
\end{eqnarray}
With the simple replacements $\widetilde{A}V\to \widetilde{V}A$ and $S\widetilde{V}$, we obtain the corresponding components $\Pi(p^{\prime2},p^2,q^2)$ for the currents $J_{-,\mu}^{\widetilde{V}A}(0)$ and $J_{-,\mu\nu}^{S\widetilde{V}}(0)$, except for the component  $\Pi_{\eta_c \omega S \widetilde{V}}(p^{\prime2},p^2,q^2)$,
\begin{eqnarray}\label{etacomegaSV}
\Pi_{\eta_c\omega S\widetilde{V}}(p^{\prime2},p^2,q^2)&=&
\frac{\lambda_{\eta_c\omega S\widetilde{V}}}{(m_Y^2-p^{\prime2})
(m_{\eta_c}^2-p^2)(m_{\omega}^2-q^2)}
+\frac{C_{\eta_c\omega S\widetilde{V}}}{(m_{\eta_c}^2-p^2)(m_{\omega}^2-q^2)}\nonumber\\
&&+\frac{\bar{\lambda}_{\eta_c\omega S\widetilde{A}}}{(m_X^2-p^{\prime2})
(m_{\eta_c}^2-p^2)(m_{\omega}^2-q^2)}+\cdots\, .
\end{eqnarray}
Where we introduce the following collective notations to simplify the formula,
\begin{eqnarray}
\lambda_{\bar{D}D\widetilde{A}V}&=&\frac{f_D^2m_D^4}{m_c^2}
\lambda_{\widetilde{A}V}G_{\bar{D}D\widetilde{A}V} \, , \nonumber \\
\lambda_{\bar{D}^*D\widetilde{A}V}&=&\frac{f_{D^*}m_{D^*}f_D m_D^2}{m_c}
\lambda_{\widetilde{A}V}G_{\bar{D}^*D\widetilde{A}V} \, ,\nonumber \\
\lambda_{\bar{D}^*D^*\widetilde{A}V}&=&f_{D^*}^2m_{D^*}^2
\lambda_{\widetilde{A}V}G_{\bar{D}^*D^*\widetilde{A}V}\, ,\nonumber \\
\lambda_{\bar{D}_0D^*\widetilde{A}V}&=&f_{D_0}m_{D_0}f_{D^*}m_{D^*}
\lambda_{\widetilde{A}V}G_{\bar{D}_0D^*\widetilde{A}V} \, ,\nonumber \\
\lambda_{\bar{D}_1D\widetilde{A}V}&=&\frac{f_{D_1}m_{D_1}f_{D}m_{D}^2}{m_c}
\lambda_{\widetilde{A}V}G_{\bar{D}_1D\widetilde{A}V} \, ,
\end{eqnarray}

\begin{eqnarray}
\lambda_{\eta_c\omega\widetilde{A}V}&=&\frac{f_{\eta_c}m_{\eta_c}^2f_{\omega}
m_{\omega}}{2m_c}
\lambda_{\widetilde{A}V}G_{\eta_c\omega\widetilde{A}V} \, ,\nonumber \\
\lambda_{J/\psi\omega\widetilde{A}V}&=&f_{J/\psi}m_{J/\psi}f_{\omega}
m_{\omega}\lambda_{\widetilde{A}V}G_{J/\psi\omega\widetilde{A}V}
\left(1+\frac{m_\omega^2}{m_Y^2}-\frac{m_{J/\psi}^2}{m_Y^2} \right)\, ,\nonumber \\
\lambda_{\chi_{c0}\omega\widetilde{A}V}&=&f_{\chi_{c0}}m_{\chi_{c0}}f_{\omega}
m_{\omega}\lambda_{\widetilde{A}V}G_{\chi_{c0}\omega\widetilde{A}V}\, ,\nonumber \\
\lambda_{\chi_{c1}\omega\widetilde{A}V}&=&f_{\chi_{c1}}m_{\chi_{c1}}f_{\omega}
m_{\omega}\lambda_{\widetilde{A}V}G_{\chi_{c1}\omega\widetilde{A}V}\, ,\nonumber \\
\lambda_{J/\psi f_0\widetilde{A}V}&=&f_{J/\psi}m_{J/\psi}f_{f_0}
m_{f_0}\lambda_{\widetilde{A}V}G_{J/\psi f_0\widetilde{A}V}\, .
\end{eqnarray}
With the simple replacements $\widetilde{A}V\to \widetilde{V}A$ and $S\widetilde{V}$, we obtain the corresponding collective notations  $\lambda$ for the currents $J_{-,\mu}^{\widetilde{V}A}(0)$ and $J_{-,\mu\nu}^{S\widetilde{V}}(0)$, except for the $\lambda_{\eta_{c} \omega S \widetilde{V}}$,  $\lambda_{J/\psi \omega S \widetilde{V}}$, $\lambda_{\chi_{c1} \omega S \widetilde{V}}$,
where
\begin{eqnarray}
\lambda_{\eta_c\omega S\widetilde{V}}&=&\frac{f_{\eta_c}m_{\eta_c}^2f_{\omega}
m_{\omega}^3}{2m_c}
\lambda_{S \widetilde{V}}G_{\eta_c\omega S \widetilde{V}} \, ,\nonumber \\
\lambda_{J/\psi\omega S\widetilde{V}}&=&f_{J/\psi}m_{J/\psi}f_{\omega}
m_{\omega}\lambda_{S\widetilde{V}}G_{J/\psi\omega S\widetilde{V}}\, , \nonumber \\
\lambda_{\chi_{c1}\omega S\widetilde{V}}&=&f_{\chi_{c1}}m_{\chi_{c1}}^3f_{\omega}
m_{\omega}\lambda_{S \widetilde{V}}G_{\chi_{c1}\omega S \widetilde{V}}\, ,
\end{eqnarray}
and
\begin{eqnarray}
\bar{\lambda}_{\eta_c\omega S\widetilde{A}}&=&\frac{f_{\eta_c}m_{\eta_c}^2f_{\omega}
m_{\omega}}{2m_c}\bar{\lambda}_{S\widetilde{A}}\bar{G}_{\eta_c\omega S\widetilde{A}}\, .
\end{eqnarray}
And we choose the standard definitions for the decay constants or pole residues,
\begin{eqnarray}
\langle 0| J^{D}(0)|D(p)\rangle&=&\frac{f_D m_D^2}{m_c}\, , \nonumber \\
\langle 0| J_\alpha^{D^*}(0)|D^*(p)\rangle&=&f_{D^*} m_{D^*}\xi_\alpha\, , \nonumber \\
\langle 0| J^{D_0}(0)|D_0(p)\rangle&=&f_{D_0} m_{D_0}\, , \nonumber \\
\langle 0| J_\alpha^{D_1}(0)|D_1(p)\rangle&=&f_{D_1} m_{D_1}\xi_\alpha\, ,
\end{eqnarray}
etc. The explicit definitions for other decay constants or pole residues are given in the Appendix.
And we use the following definitions for the hadronic coupling constants,
\begin{eqnarray}
\langle \bar{D}(p)D(q)|Y_{\widetilde{A}V}(p^\prime)\rangle&=&(p-q)\cdot \varepsilon \,G_{\bar{D}D\widetilde{A}V}\, , \nonumber\\
\langle \bar{D}(p)D(q)|Y_{\widetilde{V}A}(p^\prime)\rangle&=&(p-q)\cdot \varepsilon \,G_{\bar{D}D\widetilde{V}A}\, , \nonumber\\
\langle \bar{D}(p)D(q)|Y_{S\widetilde{V}}(p^\prime)\rangle&=&-i(p-q)\cdot \varepsilon \,G_{\bar{D}DS\widetilde{V}}\, ,
\end{eqnarray}

\begin{eqnarray}
\langle \bar{D}^*(p)D(q)|Y_{\widetilde{A}V}(p^\prime)\rangle&=&-\varepsilon^{\lambda\tau\rho\sigma}
p_\lambda \xi^*_\tau p^\prime_\rho \varepsilon_\sigma \,G_{\bar{D}^*D\widetilde{A}V}\, , \nonumber\\
\langle \bar{D}^*(p)D(q)|Y_{\widetilde{V}A}(p^\prime)\rangle&=&\varepsilon^{\lambda\tau\rho\sigma}
p_\lambda \xi^*_\tau p^\prime_\rho \varepsilon_\sigma \,G_{\bar{D}^*D\widetilde{V}A}\, , \nonumber\\
\langle \bar{D}^*(p)D(q)|Y_{S\widetilde{V}}(p^\prime)\rangle&=&-i
\varepsilon^{\lambda\tau\rho\sigma}
p_\lambda \xi^*_\tau p^\prime_\rho \varepsilon_\sigma \,G_{\bar{D}^*DS\widetilde{V}}\, ,
\end{eqnarray}
etc. The explicit definitions for other hadronic coupling constants are given in the Appendix.

In Eq.\eqref{etacomegaSV}, there are both contributions come from the $J^{PC}=1^{+-}$ and $1^{--}$ tetraquark states (see Eq.\eqref{JSV-Residue} in the Appendix, in which the $X$ state has a mass $4.01\,\rm{GeV}$ \cite{WZG-HC-spectrum-PRD}), and we cannot choose the pertinent structures to exclude the contaminations from the $J^{PC}=1^{+-}$ tetraquark state $X$, so we take it into account at the hadron side of the QCD sum rules.  The unknown parameters $C_{\bar{D}D\widetilde{A}V}$, $C_{\bar{D}^*D\widetilde{A}V}$, $C_{\bar{D}^*D^*\widetilde{A}V}$, etc, parameterize the complex interactions among the excited states in the $p^{\prime2}$ channels and the ground state conventional charmed meson pairs (or charmonium plus $\omega$), for example,
\begin{eqnarray}
C_{\bar{D}D\widetilde{A}V}&=&\int_{s_0^\prime}^{\infty} ds^\prime \frac{\rho(s^\prime,m_{\bar{D}}^2,m_D^2)}{\left(s^\prime-m_Y^2\right)\left(m_{\bar{D}}^2-p^2 \right)\left(m_D^2-q^2 \right)}\, ,
\end{eqnarray}
the spectral density $\rho(s^\prime,m_{\bar{D}}^2,m_D^2)$ is unknown, where the $s_0^\prime$ is the continuum threshold parameter for the ground state.

We choose the components $\Pi_H(p^{\prime2},p^2,q^2)$ to investigate  the hadronic coupling constants $G_{H}$, routinely, we acquire  the hadronic  spectral densities $\rho_H(s^\prime,s,u)$ through triple  dispersion relation,
\begin{eqnarray}
\Pi_{H}(p^{\prime2},p^2,q^2)&=&\int_{\Delta_s^{\prime2}}^\infty ds^{\prime} \int_{\Delta_s^2}^\infty ds \int_{\Delta_u^2}^\infty du \frac{\rho_{H}(s^\prime,s,u)}{(s^\prime-p^{\prime2})(s-p^2)(u-q^2)}\, ,
\end{eqnarray}
where the $\Delta_{s}^{\prime2}$, $\Delta_{s}^{2}$ and
$\Delta_{u}^{2}$ are thresholds, we use the notation  $H$ to represent  the components $\Pi_{\bar{D}D\widetilde{A}V}(p^{\prime2},p^2,q^2)$, $\Pi_{\bar{D}^*D\widetilde{A}V}(p^{\prime2},p^2,q^2)$, $\cdots$ at the hadron side.

At the QCD side, we accomplish  the operator product expansion up to the vacuum condensates of dimension 5 \cite{WZG-ZJX-Zc-Decay,WZG-Y4660-Decay,WZG-X4140-decay,Nielsen-decay,Azizi-decay,ChenW-decay,
WZG-X4274-decay,
WZG-X3872-decay,WZG-Zcs3985-decay,WZG-Zcs4123-decay}, then acquire the QCD spectral densities $\rho_{QCD}(p^{\prime2},s,u)$  through double dispersion relation,
\begin{eqnarray}
\Pi_{QCD}(p^{\prime2},p^2,q^2)&=& \int_{\Delta_s^2}^\infty ds \int_{\Delta_u^2}^\infty du \frac{\rho_{QCD}(p^{\prime2},s,u)}{(s-p^2)(u-q^2)}\, ,
\end{eqnarray}
as
\begin{eqnarray}
{\rm lim}_{\epsilon \to 0}\frac{{\rm Im}\,\Pi_{QCD}(s^\prime+i\epsilon,p^2,q^2)}{\pi}&=&0\, .
\end{eqnarray}
In calculations, we neglect the gluon condensates due to their tiny contributions \cite{WZG-ZJX-Zc-Decay,WZG-Y4660-Decay}.

We match the hadron side with the QCD side  bellow the continuum thresholds  to acquire   rigorous quark-hadron  duality  \cite{WZG-ZJX-Zc-Decay,WZG-Y4660-Decay},
 \begin{eqnarray}
  \int_{\Delta_s^2}^{s_{0}}ds \int_{\Delta_u^2}^{u_0}du  \frac{\rho_{QCD}(p^{\prime2},s,u)}{(s-p^2)(u-q^2)}&=& \int_{\Delta_s^2}^{s_0}ds \int_{\Delta_u^2}^{u_0}du  \left[ \int_{\Delta_{s}^{\prime2}}^{\infty}ds^\prime  \frac{\rho_H(s^\prime,s,u)}{(s^\prime-p^{\prime2})(s-p^2)(u-q^2)} \right]\, ,
\end{eqnarray}
where  the $s_0$ and $u_0$ are the continuum thresholds,  we accomplish the integral over $ds^\prime$ firstly, and introduce some unknown parameters, such as the $C_{\bar{D}D\widetilde{A}V}$, $C_{\bar{D}^*D\widetilde{A}V}$, $C_{\bar{D}^*D^*\widetilde{A}V}$, $\cdots$, to parameterize the contributions involving the higher resonances and continuum states in the $s^\prime$ channel.

We set $p^{\prime2}=p^2$ in the correlation functions $\Pi(p^{\prime 2},p^2,q^2)$, and carry out  the double Borel transform in regard  to the variables $P^2=-p^2$ and $Q^2=-q^2$ respectively, then set the Borel parameters  $T_1^2=T_2^2=T^2$  to acquire thirty QCD sum rules,
\begin{eqnarray}
&&\frac{\lambda_{\bar{D}D\widetilde{A}V}}{m_Y^2-m_D^2}
\left[\exp\left(-\frac{m_D^2}{T^2} \right)-\exp\left(-\frac{m_Y^2}{T^2} \right) \right]\exp\left(-\frac{m_D^2}{T^2}\right)+C_{\bar{D}D\widetilde{A}V}\exp\left( -\frac{m_D^2}{T^2}-\frac{m_D^2}{T^2}\right)\nonumber\\
&&=\Pi^{QCD}_{\bar{D}D\widetilde{A}V}(T^2)\, ,
\end{eqnarray}

\begin{eqnarray}
&&\frac{\lambda_{\bar{D}^*D\widetilde{A}V}}{m_Y^2-m_{D^*}^2}
\left[\exp\left(-\frac{m_{D^*}^2}{T^2} \right)-\exp\left(-\frac{m_Y^2}{T^2} \right) \right]\exp\left(-\frac{m_D^2}{T^2}\right)+C_{\bar{D}^*D\widetilde{A}V}\exp\left( -\frac{m_{D^*}^2}{T^2}-\frac{m_D^2}{T^2}\right)\nonumber\\
&&=\Pi^{QCD}_{\bar{D}^*D\widetilde{A}V}(T^2)\, ,
\end{eqnarray}

\begin{eqnarray}
&&\frac{\lambda_{\bar{D}^*D^*\widetilde{A}V}}{m_Y^2-m_{D^*}^2}
\left[\exp\left(-\frac{m_{D^*}^2}{T^2} \right)-\exp\left(-\frac{m_Y^2}{T^2} \right) \right]\exp\left(-\frac{m_{D^*}^2}{T^2}\right)+C_{\bar{D}^*D^*\widetilde{A}V}\exp\left( -\frac{m_{D^*}^2}{T^2}-\frac{m_{D^*}^2}{T^2}\right)\nonumber\\
&&=\Pi^{QCD}_{\bar{D}^*D^*\widetilde{A}V}(T^2)\, ,
\end{eqnarray}

\begin{eqnarray}
&&\frac{\lambda_{\bar{D}_0D^*\widetilde{A}V}}{m_Y^2-m_{D_0}^2}
\left[\exp\left(-\frac{m_{D_0}^2}{T^2} \right)-\exp\left(-\frac{m_Y^2}{T^2} \right) \right]\exp\left(-\frac{m_{D^*}^2}{T^2}\right)+C_{\bar{D}_0D^*\widetilde{A}V}\exp\left( -\frac{m_{D_0}^2}{T^2}-\frac{m_{D^*}^2}{T^2}\right)\nonumber\\
&&=\Pi^{QCD}_{\bar{D}_0D^*\widetilde{A}V}(T^2)\, ,
\end{eqnarray}

\begin{eqnarray}
&&\frac{\lambda_{\bar{D}_1D\widetilde{A}V}}{m_Y^2-m_{D_1}^2}
\left[\exp\left(-\frac{m_{D_1}^2}{T^2} \right)-\exp\left(-\frac{m_Y^2}{T^2} \right) \right]\exp\left(-\frac{m_{D}^2}{T^2}\right)+C_{\bar{D}_1D\widetilde{A}V}\exp\left( -\frac{m_{D_1}^2}{T^2}-\frac{m_{D}^2}{T^2}\right)\nonumber\\
&&=\Pi^{QCD}_{\bar{D}_1D\widetilde{A}V}(T^2)\, ,
\end{eqnarray}

\begin{eqnarray}
&&\frac{\lambda_{\eta_c \omega\widetilde{A}V}}{m_Y^2-m_{\eta_c}^2}
\left[\exp\left(-\frac{m_{\eta_c}^2}{T^2} \right)-\exp\left(-\frac{m_Y^2}{T^2} \right) \right]\exp\left(-\frac{m_{\omega}^2}{T^2}\right)+C_{\eta_c \omega\widetilde{A}V}\exp\left( -\frac{m_{\eta_c}^2}{T^2}-\frac{m_{\omega}^2}{T^2}\right)\nonumber\\
&&=\Pi^{QCD}_{\eta_c \omega\widetilde{A}V}(T^2)\, ,
\end{eqnarray}

\begin{eqnarray}
&&\frac{\lambda_{J/\psi \omega\widetilde{A}V}}{m_Y^2-m_{J/\psi}^2}
\left[\exp\left(-\frac{m_{J/\psi}^2}{T^2} \right)-\exp\left(-\frac{m_Y^2}{T^2} \right) \right]\exp\left(-\frac{m_{\omega}^2}{T^2}\right)+C_{J/\psi \omega\widetilde{A}V}\exp\left( -\frac{m_{J/\psi}^2}{T^2}-\frac{m_{\omega}^2}{T^2}\right)\nonumber\\
&&=\Pi^{QCD}_{J/\psi \omega\widetilde{A}V}(T^2)\, ,
\end{eqnarray}

\begin{eqnarray}
&&\frac{\lambda_{\chi_{c0} \omega\widetilde{A}V}}{m_Y^2-m_{\chi_{c0}}^2}
\left[\exp\left(-\frac{m_{\chi_{c0}}^2}{T^2} \right)-\exp\left(-\frac{m_Y^2}{T^2} \right) \right]\exp\left(-\frac{m_{\omega}^2}{T^2}\right)+C_{\chi_{c0} \omega\widetilde{A}V}\exp\left( -\frac{m_{\chi_{c0}}^2}{T^2}-\frac{m_{\omega}^2}{T^2}\right)\nonumber\\
&&=\Pi^{QCD}_{\chi_{c0} \omega\widetilde{A}V}(T^2)\, ,
\end{eqnarray}

\begin{eqnarray}
&&\frac{\lambda_{\chi_{c1} \omega\widetilde{A}V}}{m_Y^2-m_{\chi_{c1}}^2}
\left[\exp\left(-\frac{m_{\chi_{c1}}^2}{T^2} \right)-\exp\left(-\frac{m_Y^2}{T^2} \right) \right]\exp\left(-\frac{m_{\omega}^2}{T^2}\right)+C_{\chi_{c1} \omega\widetilde{A}V}\exp\left( -\frac{m_{\chi_{c1}}^2}{T^2}-\frac{m_{\omega}^2}{T^2}\right)\nonumber\\
&&=\Pi^{QCD}_{\chi_{c1} \omega\widetilde{A}V}(T^2)\, ,
\end{eqnarray}

\begin{eqnarray}
&&\frac{\lambda_{J/\psi f_0\widetilde{A}V}}{m_Y^2-m_{J/\psi}^2}
\left[\exp\left(-\frac{m_{J/\psi}^2}{T^2} \right)-\exp\left(-\frac{m_Y^2}{T^2} \right) \right]\exp\left(-\frac{m_{f_0}^2}{T^2}\right)+C_{J/\psi f_0\widetilde{A}V}\exp\left( -\frac{m_{J/\psi}^2}{T^2}-\frac{m_{f_0}^2}{T^2}\right)\nonumber\\
&&=\Pi^{QCD}_{J/\psi f_0\widetilde{A}V}(T^2)\, .
\end{eqnarray}
With the simple replacements $\widetilde{A}V\to \widetilde{V}A$ and $S\widetilde{V}$, we obtain the corresponding QCD sum rules   for the currents $J_{-,\mu}^{\widetilde{V}A}(0)$ and $J_{-,\mu\nu}^{S\widetilde{V}}(0)$, except for the $\eta_c \omega $ channel,
\begin{eqnarray}
&&\frac{\lambda_{\eta_c \omega S\widetilde{V}}}{m_Y^2-m_{\eta_c}^2}
\left[\exp\left(-\frac{m_{\eta_c}^2}{T^2} \right)-\exp\left(-\frac{m_Y^2}{T^2} \right) \right]\exp\left(-\frac{m_{\omega}^2}{T^2}\right)+C_{\eta_c \omega S\widetilde{V}}\exp\left( -\frac{m_{\eta_c}^2}{T^2}-\frac{m_{\omega}^2}{T^2}\right)\nonumber\\
&&+\frac{\bar{\lambda}_{\eta_c \omega S\widetilde{A}}}{m_X^2-m_{\eta_c}^2}
\left[\exp\left(-\frac{m_{\eta_c}^2}{T^2} \right)-\exp\left(-\frac{m_X^2}{T^2} \right) \right]\exp\left(-\frac{m_{\omega}^2}{T^2}\right)=\Pi^{QCD}_{\eta_c \omega S\widetilde{V}}(T^2)\, ,
\end{eqnarray}
where
\begin{eqnarray}
\Pi^{QCD}_{\bar{D}D\widetilde{A}V}(T^2)&=&\frac{9m_c}{128\pi^4}\int_{m_c^2}^{s^0_D}ds
\int_{m_c^2}^{s^0_D}du \left(1-\frac{m_c^2}{s}\right)^2\left(1-\frac{m_c^2}{u}\right)^2 u \, \exp\left(-\frac{s+u}{T^2} \right) \nonumber\\
&&-\frac{3m_c^2\langle\bar{q}q\rangle}{16\pi^2}\int_{m_c^2}^{s^0_D}ds \left(1-\frac{m_c^2}{s}\right)^2 \exp\left(-\frac{s+m_c^2}{T^2} \right)\nonumber\\
&&-\frac{3\langle\bar{q}q\rangle}{16\pi^2}\int_{m_c^2}^{s^0_D}du \left(1-\frac{m_c^2}{u}\right)^2 u\,\exp\left(-\frac{u+m_c^2}{T^2} \right)\nonumber\\
&&+\frac{3m_c^4\langle\bar{q}g_s\sigma G  q\rangle}{64\pi^2T^4}\int_{m_c^2}^{s^0_D}ds \left(1-\frac{m_c^2}{s}\right)^2 \exp\left(-\frac{s+m_c^2}{T^2} \right)\nonumber\\
&&+\frac{3\langle\bar{q}g_s\sigma G q\rangle}{64\pi^2T^2}\left( 1+\frac{m_c^2}{T^2}\right)\int_{m_c^2}^{s^0_D}du \left(1-\frac{m_c^2}{u}\right)^2u\, \exp\left(-\frac{u+m_c^2}{T^2} \right)\nonumber\\
&&-\frac{3m_c^2\langle\bar{q}g_s\sigma G q\rangle}{32\pi^2T^2}\int_{m_c^2}^{s^0_D}ds \left(1-\frac{m_c^2}{s}\right)^2 \exp\left(-\frac{s+m_c^2}{T^2} \right)\nonumber\\
&&-\frac{3\langle\bar{q}g_s\sigma G q\rangle}{64\pi^2T^2}\int_{m_c^2}^{s^0_D}du \left(1-\frac{m_c^2}{u}\right)^2u\, \exp\left(-\frac{u+m_c^2}{T^2} \right)\nonumber\\
&&+\frac{m_c^2\langle\bar{q}g_s\sigma G q\rangle}{64\pi^2}\int_{m_c^2}^{s^0_D}ds \frac{1}{s} \exp\left(-\frac{s+m_c^2}{T^2} \right)\nonumber\\
&&+\frac{\langle\bar{q}g_s\sigma G q\rangle}{192\pi^2}\int_{m_c^2}^{s^0_D}du \left(3-2\frac{m_c^2}{u}\right)\, \exp\left(-\frac{u+m_c^2}{T^2} \right)\, ,
\end{eqnarray}
the explicit expressions of the QCD side of other QCD sum rules are given in the Appendix.
We  take the unknown parameters $C_{\bar{D}D\widetilde{A}V}$, $C_{\bar{D}^*D\widetilde{A}V}$, $C_{\bar{D}^*D^*\widetilde{A}V}$, $\cdots$ as free parameters, and adjust the suitable values to obtain flat Borel platforms for the hadronic coupling constants  \cite{WZG-ZJX-Zc-Decay,WZG-Y4660-Decay,WZG-X4140-decay,WZG-X4274-decay,
WZG-X3872-decay,WZG-Zcs3985-decay}.

\section{Numerical results and discussions}
We take  the standard values of the  vacuum condensates
$\langle \bar{q}q \rangle=-(0.24\pm 0.01\, \rm{GeV})^3$,
$\langle\bar{q}g_s\sigma G q \rangle=m_0^2\langle \bar{q}q \rangle$ and
$m_0^2=(0.8 \pm 0.1)\,\rm{GeV}^2$     at the   energy scale  $\mu=1\, \rm{GeV}$
\cite{SVZ79,Reinders85,Colangelo-Review},  and take the $\overline{MS}$  mass $m_{c}(m_c)=(1.275\pm0.025)\,\rm{GeV}$ from the Particle Data Group \cite{PDG}. We set $m_u=m_d=0$ and take account of
the energy-scale dependence from re-normalization group equation,
\begin{eqnarray}
\langle\bar{q}q \rangle(\mu)&=&\langle\bar{q}q \rangle({\rm 1GeV})\left[\frac{\alpha_{s}({\rm 1GeV})}{\alpha_{s}(\mu)}\right]^{\frac{12}{33-2n_f}}\, , \nonumber\\
 \langle\bar{q}g_s \sigma Gq \rangle(\mu)&=&\langle\bar{q}g_s \sigma Gq \rangle({\rm 1GeV})\left[\frac{\alpha_{s}({\rm 1GeV})}{\alpha_{s}(\mu)}\right]^{\frac{2}{33-2n_f}}\, , \nonumber\\
 m_c(\mu)&=&m_c(m_c)\left[\frac{\alpha_{s}(\mu)}{\alpha_{s}(m_c)}\right]^{\frac{12}{33-2n_f}} \, ,\nonumber\\
 \alpha_s(\mu)&=&\frac{1}{b_0t}\left[1-\frac{b_1}{b_0^2}\frac{\log t}{t} +\frac{b_1^2(\log^2{t}-\log{t}-1)+b_0b_2}{b_0^4t^2}\right]\, ,
\end{eqnarray}
  where   $t=\log \frac{\mu^2}{\Lambda_{QCD}^2}$, $b_0=\frac{33-2n_f}{12\pi}$, $b_1=\frac{153-19n_f}{24\pi^2}$, $b_2=\frac{2857-\frac{5033}{9}n_f+\frac{325}{27}n_f^2}{128\pi^3}$,  $\Lambda_{QCD}=210\,\rm{MeV}$, $292\,\rm{MeV}$  and  $332\,\rm{MeV}$ for the flavors  $n_f=5$, $4$ and $3$, respectively  \cite{PDG,Narison-mix}, and we choose  $n_f=4$, and evolve all the input parameters to the energy scale  $\mu=1\,\rm{GeV}$.

 At the hadron side, we take the masses $m_{J/\psi}=3.0969\,\rm{GeV}$,
$m_{\eta_c}=2.9839\,\rm{GeV}$, $m_{\chi_{c1}}=3.51067\,\rm{GeV}$,
$m_{\chi_{c0}}=3.41471\,\rm{GeV}$, $m_{D^*}=2.00685\,\rm{GeV}$, $m_{D}=1.86484\,\rm{GeV}$, $m_\omega=0.78266\,\rm{GeV}$,
$m_{f_0}=0.550\,\rm{GeV}$ from the Particle Data Group \cite{PDG},
   $m_{D_0}=2.40\,\rm{GeV}$, $m_{D_1}=2.42\,\rm{GeV}$  \cite{WZG-heavy-decay},
   $M_{Y(\widetilde{V}A)}=4.53 \,\rm{GeV}$,
  $M_{Y(\widetilde{A}V)}=4.48\,\rm{GeV}$ and
  $M_{Y(S\widetilde{V})}=4.50 \,\rm{GeV}$ \cite{WZG-NPB-cucd-Vector} from the QCD sum rules.

 The two-body strong decays of the three $Y$ states take place through either S-wave or P-wave, and the partial decay widths are proportional to $|\vec{p}|$ or $|\vec{p}|^3$, and depend on the masses of the final-state mesons, where the $\vec{p}$    is the
  three-momentum in center-of-mass of the $Y$ states. From the Particle Data Group, we can see clearly that the uncertainties of the masses are very small except for the $f_0(500)$, we would take the central values of the masses for all the mesons in a unform manner. As for the $D_0^*(2300)$, its nature is still under debate, we take the masses of the cousins $D_0$ and $D_1$ from the QCD sum rules. Although the uncertainty of the mass of the $f_0(500)$ is rather large, however, the partial decay width is rather small, taking the central value would not lead to much uncertainties of the total widths. Accordingly, we take the central values of the decay constants and continuum threshold parameters, which are determined from the two-point correlation function QCD sum rules.

We write down the decay constants or pole residues
  $f_{D^*}=263\,\rm{MeV}$, $f_{D}=208\,\rm{MeV}$, $f_{D_0}=373\,\rm{MeV}$, $f_{D_1}=332\,\rm{MeV}$    \cite{WZG-heavy-decay},
  $f_{J/\psi}=0.418 \,\rm{GeV}$, $f_{\eta_c}=0.387 \,\rm{GeV}$  \cite{Becirevic},
  $f_{\chi_{c0}}=0.359\,\rm{GeV}$, $f_{\chi_{c1}}=0.338\,\rm{GeV}$ \cite{Charmonium-PRT},
  $f_{\rho}=0.215\,\rm{GeV}$, $f_{\omega}=f_{\rho}$ \cite{PBall-decay-Kv},
  $f_{f_0}=0.350\,\rm{GeV}$ \cite{ChengHY-2022,WZG-EPJC-scalar}, $\lambda_{\widetilde{V}A}=1.03\times 10^{-1}\, \rm{GeV}^5$,
       $\lambda_{\widetilde{A}V}=9.47\times 10^{-2}\, \rm{GeV}^5$ and
      $\lambda_{S\widetilde{V}}=4.78\times 10^{-2}\, \rm{GeV}^4/4.50$ \cite{WZG-NPB-cucd-Vector} from the QCD sum rules.

Furthermore, we take the continuum threshold parameters
    $s^0_{D^*}=6.4\,\rm{GeV}^2$, $s^0_{D}=6.2\,\rm{GeV}^2$, $s^0_{D_0}=8.3\,\rm{GeV}^2$, $s^0_{D_1}=8.6\,\rm{GeV}^2$   \cite{WZG-heavy-decay},
  $s^0_{\rho}=(1.2\,\rm{GeV})^2$, $s^0_\omega=s^0_{\rho}$  \cite{PBall-decay-Kv},
   $s^0_{f_0}=1.0\,\rm{GeV}^2$ \cite{ChengHY-2022,WZG-EPJC-scalar},
  $s^0_{J/\psi}=(3.6\,\rm{GeV})^2$, $s^0_{\eta_c}=(3.5\,\rm{GeV})^2$,
  $s^0_{\chi_{c0}}=(3.9\,\rm{GeV})^2$, $s^0_{\chi_{c1}}=(4.0\,\rm{GeV})^2$ from the two-point QCD sum rules combined with the experimental data \cite{PDG,Becirevic,Charmonium-PRT}. On the one hand, we can reproduce central values of the experimental masses, on the other hand, we can exclude the contaminations from the higher resonances and continuum states. In calculations, we only take account of the uncertainties of the quark masses and vacuum condensates, which are absorbed into the decay constants, pole residues  and hadronic coupling constants reasonably to avoid doubly counting.

Generally speaking, we expect that the two-point and three-point correlation function QCD sum rules have the same Borel parameters, as the same masses and decay constants are involved. In practical calculations, different QCD sum rules have different Borel parameters. In the two-point correlation function QCD sum rules, the quarks in the current $J_A(x)$ are contracted with the antiquarks in the current $J^\dagger_A(0)$ or $\bar{J}_A(0)$, while in the three-point correlation function QCD sum rules, the quarks in the current $J_A(x)$ are contracted with the antiquarks in the current $J_B(y)$ or $J^\dagger_C(0)$ or $\bar{J}_C(0)$, the spectral representations are quite different, it is not odd that the Borel parameters are also different. We would obtain flat platforms, which minimize the uncertainties.
In calculations, we fit the free parameters as
\begin{eqnarray}
C_{\bar{D}D\widetilde{A}V}&=&0.00045\,{\rm GeV^5}\times T^2\, , \nonumber\\
C_{\bar{D}^*D\widetilde{A}V}&=&-0.000003\,{\rm GeV^4}\times T^2\, , \nonumber\\
C_{\bar{D}^*D^*\widetilde{A}V}&=&0.0001\,{\rm GeV^5}\times T^2\, , \nonumber\\
C_{\bar{D}_0D^*\widetilde{A}V}&=&0.000055\,{\rm GeV^4}\times T^2\, , \nonumber\\
C_{\bar{D}_1D\widetilde{A}V}&=&0.0031\,{\rm GeV^6}\times T^2\, , \nonumber\\
C_{\eta_c\omega\widetilde{A}V}&=&-0.000082\,{\rm GeV^4}\times T^2\, , \nonumber\\
C_{J/\psi\omega\widetilde{A}V}&=&0.0\, , \nonumber\\
C_{\chi_{c0}\omega\widetilde{A}V}&=&0.00085\,{\rm GeV^6}\times T^2\, , \nonumber\\
C_{\chi_{c1}\omega\widetilde{A}V}&=&-0.000019\,{\rm GeV^3}\times T^2\, , \nonumber\\
C_{J/\psi f_0\widetilde{A}V}&=&0.00085\,{\rm GeV^6}\times T^2\, ,
\end{eqnarray}
\begin{eqnarray}
C_{\bar{D}D \widetilde{V}A}&=&0.0000015\,{\rm GeV^5}\times T^2\, , \nonumber\\
C_{\bar{D}^*D\widetilde{V}A}&=&0.000038\,{\rm GeV^4}\times T^2\, , \nonumber\\
C_{\bar{D}^*D^*\widetilde{V}A}&=&0.0000054\,{\rm GeV^5}\times T^2\, , \nonumber\\
C_{\bar{D}_0D^*\widetilde{V}A}&=&0.00264\,{\rm GeV^6}\times T^2\, , \nonumber\\
C_{\bar{D}_1D\widetilde{V}A}&=&0.000055\,{\rm GeV^4}\times T^2\, , \nonumber\\
C_{\eta_c\omega \widetilde{V}A}&=&-0.00006\,{\rm GeV^4}\times T^2\, , \nonumber\\
C_{J/\omega\omega \widetilde{V}A}&=&0.0\, , \nonumber\\
C_{\chi_{c0}\omega \widetilde{V}A}&=&0.00087\,{\rm GeV^6}\times T^2\, , \nonumber\\
C_{\chi_{c1}\omega \widetilde{V}A}&=&-0.000018\,{\rm GeV^3}\times T^2\, , \nonumber\\
C_{J/\psi f_0 \widetilde{V}A}&=&0.00075\,{\rm GeV^6}\times T^2\, ,
\end{eqnarray}
\begin{eqnarray}
C_{\bar{D}D S\widetilde{V}}&=&0.00014\, {\rm GeV^6}+0.00002\,{\rm GeV^4}\times T^2\, ,\nonumber\\
C_{\bar{D}^*D S\widetilde{V}}&=&0.0000006\,{\rm GeV^3}\times T^2\, , \nonumber\\
C_{\bar{D}^*D^* S\widetilde{V}}&=&-0.000002\,{\rm GeV^4}\times T^2\, , \nonumber\\
C_{\bar{D}_0D^* S\widetilde{V}}&=&0.0012\, {\rm GeV^7}+0.000164\,{\rm GeV^5}\times T^2\, , \nonumber\\
C_{\bar{D}_1D S\widetilde{V}}&=&0.0017\, {\rm GeV^7}+0.000205\,{\rm GeV^5}\times T^2\, , \nonumber\\
C_{\eta_c \omega S\widetilde{V}}&=&0.00014\, {\rm GeV^7}\, , \nonumber\\
\bar{\lambda}_{\eta_c\omega S\widetilde{A}}&=&0.0012\, {\rm GeV^7}\times T^2\, , \nonumber\\
C_{J/\psi\omega S\widetilde{V}}&=&-0.0005\, {\rm GeV^6}-0.0000216\,{\rm GeV^4}\times T^2\, , \nonumber\\
C_{\chi_{c0}\omega S\widetilde{V}}&=&-0.0018\, {\rm GeV^7}-0.000156\,{\rm GeV^5}\times T^2\, , \nonumber\\
C_{\chi_{c1}\omega S\widetilde{V}}&=&0.0012\, {\rm GeV^8}+0.00015\,{\rm GeV^6}\times T^2\, , \nonumber\\
C_{J/\psi f_0 S\widetilde{V}}&=&0.0005\, {\rm GeV^7}+0.00006\,{\rm GeV^5}\times T^2\, ,
\end{eqnarray}
to obtain the Borel windows, which are shown explicitly in Table \ref{BorelP}.
We obtain uniform flat platforms  $T^2_{max}-T^2_{min}=1\,\rm{GeV}^2$, where the max and min denote the maximum and minimum, respectively.

\begin{figure}
\centering
\includegraphics[totalheight=5cm,width=7cm]{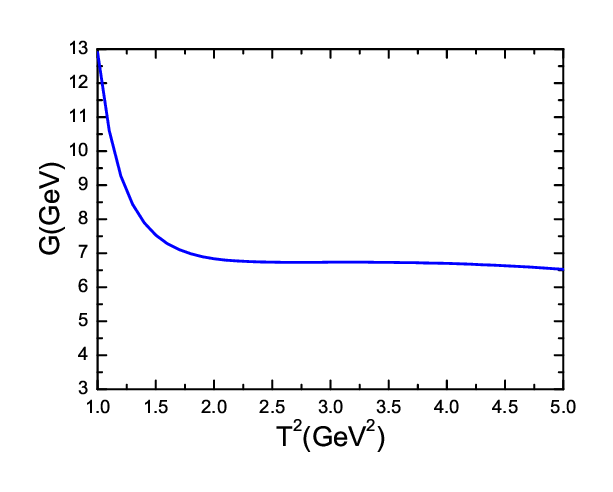}
\caption{The hadron-coupling constant $G_{\bar{D}_1 D \widetilde{A}V}$   via the  Borel  parameter.}\label{hadron-coupling-fig}
\end{figure}

In Fig.\ref{hadron-coupling-fig}, we plot the hadronic coupling constant $G_{\bar{D}_1 D \widetilde{A} V}$ with variation of the Borel parameter at large interval as an example. In the Borel windows, there appear very flat platforms  indeed, now we would extract the hadron coupling constants.

We estimate the uncertainties of the hadronic  coupling constants routinely. For an input parameter $\xi$, $\xi= \bar{\xi} +\delta \xi$,  the left side of the QCD sum rules can be written as  $\lambda_{\widetilde{A}V}f_{\bar{D}}f_{D}G_{\bar{D}D\widetilde{A}V} = \bar{\lambda}_{\widetilde{A}V}\bar{f}_{\bar{D}}\bar{f}_{D}\bar{G}_{\bar{D}D\widetilde{A}V}
+\delta\,\lambda_{\widetilde{A}V}f_{\bar{D}}f_{D}G_{\bar{D}D\widetilde{A}V}$, $C_{\bar{D}D\widetilde{A}V} = \bar{C}_{\bar{D}D\widetilde{A}V}+\delta C_{\bar{D}D\widetilde{A}V}$, $\cdots$, where
\begin{eqnarray}\label{Uncertainty-4}
\delta\,\lambda_{\widetilde{A}V}f_{\bar{D}}f_{D}G_{\bar{D}D\widetilde{A}V} &=&\bar{\lambda}_{\widetilde{A}V}\bar{f}_{\bar{D}}\bar{f}_{D}\bar{G}_{\bar{D}D\widetilde{A}V}
\left( \frac{\delta f_{\bar{D}}}{\bar{f}_{\bar{D}}} +\frac{\delta f_{D}}{\bar{f}_{D}}+\frac{\delta \lambda_{\widetilde{A}V}}{\bar{\lambda}_{\widetilde{A}V}}
+\frac{\delta G_{\bar{D}D\widetilde{A}V}}{\bar{G}_{\bar{D}D\widetilde{A}V}}\right)\, ,
\end{eqnarray}
$\cdots$.
We can set $\delta C_{\bar{D}D\widetilde{A}V}=0$,  $\frac{\delta f_{\bar{D}}}{\bar{f}_{\bar{D}}} =\frac{\delta f_{D}}{\bar{f}_{D}}=\frac{\delta \lambda_{\widetilde{A}V}}{\bar{\lambda}_{\widetilde{A}V}}
=\frac{\delta G_{\bar{D}D\widetilde{A}V}}{\bar{G}_{\bar{D}D\widetilde{A}V}}$, $\cdots$ approximately.

After taking into account the uncertainties, we obtain the values of the hadronic coupling constants, which are shown explicitly in Table \ref{BorelP}.
Then we obtain the partial decay widths routinely, and show them explicitly in Table \ref{Width-Part}.

\begin{table}
\begin{center}
\begin{tabular}{|c|c|c|c|c|c|c|c|c|}\hline\hline
Channels                         &$T^2(\rm{GeV}^2)$ &$G $    \\ \hline

$\bar{D}D\widetilde{A}V$         &$3.8-4.8$  &$2.11\pm0.10$   \\

$\bar{D}^*D\widetilde{A}V$       &$4.7-5.7$  &$(4.49\pm0.15)\times\rm{10^{-2} \,GeV^{-1}}$ \\

$\bar{D}^*D^*\widetilde{A}V$     &$4.1-5.1$  &$0.95\pm0.04$      \\

$\bar{D}_0D^*\widetilde{A}V$     &$4.4-5.4$  &$0.30\pm0.01\,\rm{GeV^{-1}}$        \\

$\bar{D}_1D\widetilde{A}V$       &$2.5-3.5$  &$6.73\pm0.31\,\rm{GeV}$   \\

$\eta_c\omega\widetilde{A}V$     &$3.7-4.7$  &$-(0.35\pm0.03)\,\rm{GeV^{-1}}$     \\

$J/\psi\omega\widetilde{A}V$     &$---$      &$0.0$     \\

$\chi_{c0}\omega\widetilde{A}V$  &$3.8-4.8$  &$2.72\pm0.20\,\rm{GeV}$     \\

$\chi_{c1}\omega\widetilde{A}V$  &$5.2-6.2$  &$-(0.25\pm0.01)\,\rm{GeV}^{-2}$     \\

$J/\psi f_0\widetilde{A}V$       &$3.8-4.8$  &$2.51\pm0.15\,\rm{GeV}$     \\ \hline

$\bar{D}D \widetilde{V}A$        &$4.4-5.4$  &$-(4.02\pm0.16)\times\rm{10^{-2}}$ \\

$\bar{D}^*D\widetilde{V}A$       &$4.0-5.0$  &$0.30\pm0.01\,\rm{ GeV^{-1}}$ \\

$\bar{D}^*D^* \widetilde{V}A$    &$4.6-5.6$  &$-(6.00\pm0.20)\times\rm{10^{-2}}$ \\

$\bar{D}_0D^*\widetilde{V}A$     &$2.6-3.6$  &$8.00\pm0.37\,\rm{ GeV}$ \\

$\bar{D}_1D\widetilde{V}A$       &$2.8-3.8$  &$0.30\pm0.01\,\rm{ GeV}^{-1}$ \\

$\eta_c\omega \widetilde{V}A$    &$5.3-6.3$  &$-(0.40\pm0.03)\,\rm{GeV^{-1}}$     \\

$J/\psi\omega \widetilde{V}A$    &$---$      &$0.0$     \\

$\chi_{c0}\omega \widetilde{V}A$ &$3.0-4.0$  &$2.56\pm0.19\,\rm{GeV}$     \\

$\chi_{c1}\omega \widetilde{V}A$ &$5.4-6.4$  &$-(0.24\pm0.01)\,\rm{GeV}^{-2}$     \\

$J/\psi f_0 \widetilde{V}A$      &$4.6-5.6$  &$2.60\pm0.15\,\rm{GeV}$     \\ \hline

$\bar{D}D S\widetilde{V}$        &$2.5-3.5$  &$0.60\pm0.06 $     \\

$\bar{D}^*D S\widetilde{V}$      &$4.7-5.7$  &$-(7.00\pm0.24)\times 10^{-2}\,\rm{GeV}^{-1} $   \\

$\bar{D}^*D^* S\widetilde{V}$    &$4.0-5.0$  &$0.18\pm0.01$     \\

$\bar{D}_0D^* S\widetilde{V}$    &$3.3-4.3$  &$2.23\pm0.24\,\rm{GeV} $   \\

$\bar{D}_1D S\widetilde{V}$      &$3.5-4.5$  &$4.21\pm0.37\,\rm{GeV} $   \\

$\eta_c\omega S\widetilde{V}$    &$2.8-3.8$  &$1.71\pm0.31\,\rm{GeV}^{-1} $   \\

$J/\psi\omega S\widetilde{V}$    &$3.7-4.7$  &$-(1.08\pm0.19) $   \\

$\chi_{c0}\omega S\widetilde{V}$ &$4.4-5.4$  &$-(5.56\pm0.97)\,\rm{GeV} $   \\

$\chi_{c1}\omega S\widetilde{V}$ &$3.1-4.1$  &$0.29\pm0.04 $   \\

$J/\psi f_0 S\widetilde{V}$      &$3.5-4.5$  &$0.47\pm0.14\,\rm{GeV} $   \\

\hline\hline
\end{tabular}
\end{center}
\caption{ The Borel parameters and hadronic coupling constants. }\label{BorelP}
\end{table}

\begin{table}
\begin{center}
\begin{tabular}{|c|c|c|c|c|c|c|c|c|}\hline\hline
Channels                                               &$\Gamma(\rm{MeV})$ \\ \hline

$Y_{\widetilde{A}V}\to \bar{D}^0D^0$, $\bar{D}^-D^+$   &$22.5\pm2.1$      \\

$Y_{\widetilde{A}V}\to \frac{\bar{D}^{0*}D^0-\bar{D}^{0}D^{*0}}{\sqrt{2}}$, $\frac{\bar{D}^{-*}D^+-\bar{D}^{-}D^{*+}}{\sqrt{2}}$   &$0.08\pm0.01 $ \\

$Y_{\widetilde{A}V}\to \bar{D}^{*0}D^{*0}$, $\bar{D}^{*-}D^{*+}$  &$9.93\pm0.84$      \\

$Y_{\widetilde{A}V}\to\frac{\bar{D}^0_0D^{*0}-\bar{D}^{*0}D^{0}_0}{\sqrt{2}}$, $\frac{\bar{D}^-_0D^{*+}-\bar{D}^{*-}D^{+}_0}{\sqrt{2}}$     &$1.92\pm0.13$        \\

$Y_{\widetilde{A}V}\to\frac{\bar{D}^0_1D^{0}-\bar{D}^{0}D^{0}_1}{\sqrt{2}}$, $\frac{\bar{D}^-_1D^{+}-\bar{D}^{-}D^{+}_1}{\sqrt{2}}$  &$59.7\pm5.5$     \\

$Y_{\widetilde{A}V}\to\eta_c\omega$     &$3.83\pm0.66$     \\

$Y_{\widetilde{A}V}\to J/\psi\omega$    &$0.0$     \\

$Y_{\widetilde{A}V}\to\chi_{c0}\omega$  &$11.3\pm1.7$     \\

$Y_{\widetilde{A}V}\to\chi_{c1}\omega$  &$24.4\pm1.9$     \\

$Y_{\widetilde{A}V}\to J/\psi f_0(500)$ &$13.9\pm1.7$     \\ \hline

$Y_{\widetilde{V}A}\to \bar{D}^0D^0$, $\bar{D}^-D^+$   &$0.009\pm0.001$ \\

$Y_{\widetilde{V}A}\to \frac{\bar{D}^{0*}D^0-\bar{D}^{0}D^{*0}}{\sqrt{2}}$, $\frac{\bar{D}^{-*}D^+-\bar{D}^{-}D^{*+}}{\sqrt{2}}$   &$3.88\pm0.26$ \\

$Y_{\widetilde{V}A}\to \bar{D}^{*0}D^{*0}$, $\bar{D}^{*-}D^{*+}$  &$0.047\pm0.003$ \\

$Y_{\widetilde{V}A}\to\frac{\bar{D}^0_0D^{*0}-\bar{D}^{*0}D^{0}_0}{\sqrt{2}}$, $\frac{\bar{D}^-_0D^{*+}-\bar{D}^{*-}D^{+}_0}{\sqrt{2}}$  &$66.3\pm6.1$ \\

$Y_{\widetilde{V}A}\to\frac{\bar{D}^0_1D^{0}-\bar{D}^{0}D^{0}_1}{\sqrt{2}}$, $\frac{\bar{D}^-_1D^{+}-\bar{D}^{-}D^{+}_1}{\sqrt{2}}$   &$4.10\pm0.27$ \\

$Y_{\widetilde{V}A}\to\eta_c\omega $                     &$5.65\pm0.85$     \\

$Y_{\widetilde{V}A}\to J/\psi\omega $                    &$0.0$     \\

$Y_{\widetilde{V}A}\to\chi_{c0}\omega$                   &$11.1\pm1.6$     \\

$Y_{\widetilde{V}A}\to\chi_{c1}\omega $                  &$29.9\pm2.5$     \\

$Y_{\widetilde{V}A}\to J/\psi f_0(500)$                  &$15.2\pm1.8$     \\ \hline

$Y_{S\widetilde{V}}\to \bar{D}^0D^0$, $\bar{D}^-D^+$     &$1.88\pm0.38 $     \\

$Y_{S\widetilde{V}}\to \frac{\bar{D}^{0*}D^0-\bar{D}^{0}D^{*0}}{\sqrt{2}}$, $\frac{\bar{D}^{-*}D^+-\bar{D}^{-}D^{*+}}{\sqrt{2}}$  &$0.20\pm0.01 $   \\

$Y_{S\widetilde{V}}\to \frac{\bar{D}^{0*}D^0-\bar{D}^{0}D^{*0}}{\sqrt{2}}$, $\frac{\bar{D}^{-*}D^+-\bar{D}^{-}D^{*+}}{\sqrt{2}}$  &$0.38\pm0.04 $     \\

$Y_{S\widetilde{V}}\to\frac{\bar{D}^0_0D^{*0}-\bar{D}^{*0}D^{0}_0}{\sqrt{2}}$, $\frac{\bar{D}^-_0D^{*+}-\bar{D}^{*-}D^{+}_0}{\sqrt{2}}$   &$4.51\pm0.97 $   \\

$Y_{S\widetilde{V}}\to\frac{\bar{D}^0_1D^{0}-\bar{D}^{0}D^{0}_1}{\sqrt{2}}$, $\frac{\bar{D}^-_1D^{+}-\bar{D}^{-}D^{+}_1}{\sqrt{2}}$  &$24.4\pm4.3$   \\

$Y_{S\widetilde{V}}\to\eta_c\omega $      &$96.0\pm34.8 $   \\

$Y_{S\widetilde{V}}\to J/\psi\omega $     &$18.0\pm6.3 $   \\

$Y_{S\widetilde{V}}\to\chi_{c0}\omega $   &$49.4\pm17.2 $   \\

$Y_{S\widetilde{V}}\to \chi_{c1}\omega$   &$2.76\pm0.76 $   \\

$Y_{S\widetilde{V}}\to J/\psi f_0(500)$   &$0.49\pm0.29 $   \\

\hline\hline
\end{tabular}
\end{center}
\caption{ The partial decay widths. }\label{Width-Part}
\end{table}

Finally, we saturate the total widths with the summary of the partial decay widths,
\begin{eqnarray}\label{Widths}
\Gamma\left(Y_{\widetilde{A}V}\right)&=&241.6\pm 9.0\, \rm{MeV}\, , \nonumber \\
\Gamma\left(Y_{\widetilde{V}A}\right)&=&210.6\pm 9.4\, \rm{MeV}\, , \nonumber \\
\Gamma\left(Y_{S\widetilde{V}}\right)&=&229.4\pm 39.9\, \rm{MeV}\, .
\end{eqnarray}
The widths of the $Y(4484)$, $Y(4469)$ and $Y(4544)$
are $111.1\pm30.1\pm15.2\,\rm{MeV}$,
 $246.3\pm36.7\pm9.4\,{\rm MeV}$ and $116.1\pm33.5\pm1.7\, \rm{MeV}$, respectively,  from the BESIII collaboration \cite{BESIII-Y4484-CPC,BESIII-Y4469-PRL,BESIII-Y4544-2401}, which are compatible with the present calculations in magnitude. Generally speaking, the partial decay widths shown in Table \ref{Width-Part} are not independent, and have some correlations with each other, as the masses of the relevant mesons could affect the phase-space differently in different channels. As a first step screen, we take the central values of the hadron masses and estimate the uncertainties of the partial decay widths using $\delta \Gamma \propto \delta G_H^2=2 G_H \delta G_H$, then we take  square root of the sum of their squares, which maybe underestimate the uncertainties of the total widths. In addition, we have set $\frac{\delta f_{\bar{D}}}{\bar{f}_{\bar{D}}} =\frac{\delta f_{D}}{\bar{f}_{D}}=\frac{\delta \lambda_{\widetilde{A}V}}{\bar{\lambda}_{\widetilde{A}V}}
=\frac{\delta G_{\bar{D}D\widetilde{A}V}}{\bar{G}_{\bar{D}D\widetilde{A}V}}$, $\cdots$ approximately in calculating the hadronic coupling constants $G_H$.
Alternatively, if we set $\frac{\delta f_{\bar{D}}}{\bar{f}_{\bar{D}}} =\frac{\delta f_{D}}{\bar{f}_{D}}=\frac{\delta \lambda_{\widetilde{A}V}}{\bar{\lambda}_{\widetilde{A}V}}
=0$, $\cdots$ approximately, and attribute all the uncertainties originating  from the input parameters to the hadronic coupling constants $\frac{\delta G_H}{G_H}$, we obtain,
\begin{eqnarray}\label{Widths}
\Gamma\left(Y_{\widetilde{A}V}\right)&=&241.6\pm 36.0\, \rm{MeV}\, , \nonumber \\
\Gamma\left(Y_{\widetilde{V}A}\right)&=&210.6\pm 37.6\, \rm{MeV}\, , \nonumber \\
\Gamma\left(Y_{S\widetilde{V}}\right)&=&229.4\pm 159.6\, \rm{MeV}\, ,
\end{eqnarray}
which overestimate the uncertainties.

From Table \ref{Width-Part}, we can obtain the typical decay modes. For the $Y_{\widetilde{A}V}$ state, the decays,
\begin{eqnarray}
Y_{\widetilde{A}V}&\to&\frac{\bar{D}^0_1D^{0}-\bar{D}^{0}D^{0}_1}{\sqrt{2}}\, , \,\frac{\bar{D}^-_1D^{+}-\bar{D}^{-}D^{+}_1}{\sqrt{2}}\, ,
\end{eqnarray}
have the largest partial decay width $59.7\pm5.5\,\rm{MeV}$; while the decay,
\begin{eqnarray}
Y_{\widetilde{A}V} &\to& J/\psi\omega\, ,
\end{eqnarray}
has zero partial decay width.
For the $Y_{\widetilde{V}A}$ state, the decays,
\begin{eqnarray}
Y_{\widetilde{V}A}&\to&\frac{\bar{D}^0_0D^{*0}-\bar{D}^{*0}D^{0}_0}{\sqrt{2}}\, , \, \frac{\bar{D}^-_0D^{*+}-\bar{D}^{*-}D^{+}_0}{\sqrt{2}}\, ,
\end{eqnarray}
have the largest partial decay width $66.3\pm6.1\,\rm{MeV}$; while the decay,
\begin{eqnarray}
Y_{\widetilde{V}A} &\to& J/\psi\omega\, ,
\end{eqnarray}
has zero partial decay width.
For the $Y_{S\widetilde{V}}$ state, the decay,
\begin{eqnarray}
Y_{S\widetilde{V}}&\to&\eta_c\omega\, ,
\end{eqnarray}
has the largest partial decay width $96.0\pm34.8 \,\rm{MeV}$; while the decay,
\begin{eqnarray}
Y_{S\widetilde{V}} &\to& J/\psi\omega\, ,
\end{eqnarray}
has the partial decay width $18.0\pm6.3\,\rm{MeV}$. We can search for the $Y(4500)$ in  those typical decays to diagnose the nature of the $Y$ states.

\section{Conclusion}
 In the present work, we suppose that there exist three vector hidden-charm tetraquark states with the quantum numbers $J^{PC}=1^{--}$ at the energy about $4.5\,\rm{GeV}$, and investigate the two-body strong decays systematically. We carry out the operator product expansion up to the vacuum condensates of dimension  5 and neglect the small gluon condensate contributions, then we acquire thirty QCD sum rules for the hadronic coupling constants based on the rigorous quark-hadron duality proposed in our previous work. Finally we obtain the partial decay widths, and therefore the total decay widths as a summary approximately, which are about $200\,\rm{MeV}$ and are compatible with the experimental data of the $Y(4500)$ from the BESIII collaboration. The $Y(4500)$ may be one vector tetraquark state having three main Fock components, or consists  of three vector tetraquark states $Y(4484)$, $Y(4469)$ and $Y(4544)$. By searching for the typical decay modes
 $ Y \to \frac{\bar{D}^0_1D^{0}-\bar{D}^{0}D^{0}_1}{\sqrt{2}}$,
 $\frac{\bar{D}^-_1D^{+}-\bar{D}^{-}D^{+}_1}{\sqrt{2}}$, $\frac{\bar{D}^0_0D^{*0}-\bar{D}^{*0}D^{0}_0}{\sqrt{2}}$,
$ \frac{\bar{D}^-_0D^{*+}-\bar{D}^{*-}D^{+}_0}{\sqrt{2}}$,
$\eta_c\omega$, $J/\psi\omega$, we can diagnose the nature of the $Y(4500)$.

\section*{Appendix}
The explicit expressions for other ground state contributions to the correlation functions,
\begin{eqnarray}
\Pi^{\bar{D}D\widetilde{V}A}_{\mu}(p,q)&=&\Pi_{\bar{D}D\widetilde{V}A}(p^{\prime2},p^2,q^2)
\,i\left(p-q\right)_\mu+\cdots\, ,
\end{eqnarray}

\begin{eqnarray}
\Pi^{\bar{D}^*D\widetilde{V}A}_{\alpha\mu}(p,q)&=&
\Pi_{\bar{D}^*D\widetilde{V}A}(p^{\prime2},p^2,q^2)
\,\left(-i\varepsilon_{\alpha\mu\lambda\tau}p^\lambda q^\tau\right)+\cdots\, ,
\end{eqnarray}

\begin{eqnarray}
\Pi^{\bar{D}^*D^*\widetilde{V}A}_{\alpha\beta\mu}(p,q)&=&
\Pi_{\bar{D}^*D^*\widetilde{V}A}(p^{\prime2},p^2,q^2)
\,\left(-ig_{\alpha\beta}p_\mu\right)+\cdots\, ,
\end{eqnarray}

\begin{eqnarray}
\Pi^{\bar{D}_0D^*\widetilde{V}A}_{\alpha\mu}(p,q)&=&
\Pi_{\bar{D}_0D^*\widetilde{V}A}(p^{\prime2},p^2,q^2)
\,\left(-ig_{\alpha\mu}\right)+\cdots\, ,
\end{eqnarray}

\begin{eqnarray}
\Pi^{\bar{D}_1D\widetilde{V}A}_{\alpha\mu}(p,q)&=&
\Pi_{\bar{D}_1D\widetilde{V}A}(p^{\prime2},p^2,q^2)
\,\left(g_{\alpha\mu}p \cdot q\right)+\cdots\, ,
\end{eqnarray}

\begin{eqnarray}
\Pi^{\eta_c\omega\widetilde{V}A}_{\alpha\mu}(p,q)&=&
\Pi_{\eta_c\omega\widetilde{V}A}(p^{\prime2},p^2,q^2)
\,\left(i\varepsilon_{\alpha\mu\lambda\tau}p^\lambda q^\tau\right)+\cdots\, ,
\end{eqnarray}

\begin{eqnarray}
\Pi^{J/\psi\omega\widetilde{V}A}_{\alpha\beta\mu}(p,q)&=&
\Pi_{J/\psi\omega\widetilde{V}A}(p^{\prime2},p^2,q^2)
\,\left(ig_{\alpha\beta}p_\mu\right)+\cdots\, ,
\end{eqnarray}

\begin{eqnarray}
\Pi^{\chi_{c0}\omega\widetilde{V}A}_{\alpha\mu}(p,q)&=&
\Pi_{\chi_{c0}\omega\widetilde{V}A}(p^{\prime2},p^2,q^2)
\,\left(-ig_{\alpha\mu}\right)+\cdots\, ,
\end{eqnarray}

\begin{eqnarray}
\Pi^{\chi_{c1}\omega\widetilde{V}A}_{\alpha\beta\mu}(p,q)&=&
\Pi_{\chi_{c1}\omega\widetilde{V}A}(p^{\prime2},p^2,q^2)
\,\left(\varepsilon_{\alpha\beta\mu\tau}p^\tau \,p \cdot q\right)+\cdots\, ,
\end{eqnarray}

\begin{eqnarray}
\Pi^{J/\psi f_0\widetilde{V}A}_{\alpha\mu}(p,q)&=&
\Pi_{J/\psi f_0\widetilde{V}A}(p^{\prime2},p^2,q^2)
\,\left(-ig_{\alpha\mu}\right)+\cdots\, ,
\end{eqnarray}

\begin{eqnarray}
\Pi^{\bar{D}D S\widetilde{V}}_{\mu\nu}(p,q)&=&\Pi_{\bar{D}DS\widetilde{V}}(p^{\prime2},p^2,q^2)
\,\left(2\varepsilon_{\mu\nu\lambda\tau}p^\lambda q^\tau \right)+\cdots\, ,
\end{eqnarray}

\begin{eqnarray}
\Pi^{\bar{D}^* D S\widetilde{V}}_{\alpha\mu\nu}(p,q)&=&
\Pi_{\bar{D}^* D S\widetilde{V}}(p^{\prime2},p^2,q^2)
\,\,p \cdot q \,\left[ g_{\mu\alpha}(p-q)_\nu-g_{\nu\alpha}(p-q)_\mu\right]+\cdots\, ,
\end{eqnarray}

\begin{eqnarray}
\Pi^{\bar{D}^*D^*S\widetilde{V}}_{\alpha\beta\mu\nu}(p,q)&=&
\Pi_{\bar{D}^*D^*S\widetilde{V}}(p^{\prime2},p^2,q^2)
\,\left(-2g_{\alpha\beta}\varepsilon_{\mu\nu\lambda\tau} p^\lambda q^\tau \right)+\cdots\, ,
\end{eqnarray}

\begin{eqnarray}
\Pi^{\bar{D}_0D^*S\widetilde{V}}_{\alpha\mu\nu}(p,q)&=&
\Pi_{\bar{D}_0D^*S\widetilde{V}}(p^{\prime2},p^2,q^2)
\,\left(-\varepsilon_{\alpha\mu\nu\tau}p^\tau \right)+\cdots\, ,
\end{eqnarray}

\begin{eqnarray}
\Pi^{\bar{D}_1DS\widetilde{V}}_{\alpha\mu\nu}(p,q)&=&
\Pi_{\bar{D}_1DS\widetilde{V}}(p^{\prime2},p^2,q^2)
\,\left(-i\varepsilon_{\alpha\mu\nu\tau}p^\tau \right)+\cdots\, ,
\end{eqnarray}

\begin{eqnarray}
\Pi^{\eta_c\omega S\widetilde{V}}_{\alpha\mu\nu}(p,q)&=&
\Pi_{\eta_c\omega S\widetilde{V}}(p^{\prime2},p^2,q^2)
\,\left(g_{\nu\alpha}p_\mu-g_{\mu\alpha}p_\nu \right)+\cdots\, ,
\end{eqnarray}

\begin{eqnarray}
\Pi^{J/\psi\omega S\widetilde{V}}_{\alpha\beta\mu\nu}(p,q)&=&
\Pi_{J/\psi\omega S\widetilde{V}}(p^{\prime2},p^2,q^2)
\,\left(-2g_{\alpha\beta}\varepsilon_{\mu\nu\lambda\tau }p^\lambda q^\tau\right)+\cdots\, ,
\end{eqnarray}

\begin{eqnarray}
\Pi^{\chi_{c0}\omega S\widetilde{V}}_{\alpha\mu\nu}(p,q)&=&
\Pi_{\chi_{c0}\omega S\widetilde{V}}(p^{\prime2},p^2,q^2)
\,\left(-\varepsilon_{\alpha\mu\nu\tau}q^\tau\right)+\cdots\, ,
\end{eqnarray}

\begin{eqnarray}
\Pi^{\chi_{c1}\omega S\widetilde{V}}_{\alpha\beta\mu\nu}(p,q)&=&
\Pi_{\chi_{c1}\omega S\widetilde{V}}(p^{\prime2},p^2,q^2)
\, i  \left(g_{\alpha\nu}g_{\beta\mu}-g_{\alpha\mu}g_{\beta\nu} \right)+\cdots\, ,
\end{eqnarray}

\begin{eqnarray}
\Pi^{J/\psi f_0 S\widetilde{V}}_{\alpha\mu\nu}(p,q)&=&
\Pi_{J/\psi f_0 S\widetilde{V}}(p^{\prime2},p^2,q^2)
\,\left(-\varepsilon_{\alpha\mu\nu\tau}p^\tau\right)+\cdots\, .
\end{eqnarray}

The explicit definitions for other decay constants or pole residues,
\begin{eqnarray}
\langle 0| J^{\eta_c}(0)|\eta_c(p)\rangle&=&\frac{f_{\eta_c} m_{\eta_c}^2}{2m_c}\, ,\nonumber \\
\langle 0| J_\alpha^{J/\psi}(0)|J/\psi(p)\rangle&=&f_{J/\psi} m_{J/\psi}\xi_\alpha\, , \nonumber \\
\langle 0| J^{\chi_{c0}}(0)|\chi_{c0}(p)\rangle&=&f_{\chi_{c0}} m_{\chi_{c0}}\, , \nonumber \\
\langle 0| J_\alpha^{\chi_{c1}}(0)|\chi_{c1}(p)\rangle&=&f_{\chi_{c1}} m_{\chi_{c1}}\xi_\alpha\, ,
\end{eqnarray}

\begin{eqnarray}
\langle 0| J_\alpha^{\omega}(0)|\omega(q)\rangle&=&f_{\omega} m_{\omega}\xi_\alpha\, , \nonumber \\
\langle 0| J^{f_0}(0)|f_0(q)\rangle&=&f_{f_0} m_{f_0}\, ,
\end{eqnarray}

\begin{eqnarray}
\langle 0| J_{-,\mu}^{\widetilde{A}V}(0)|Y_{\widetilde{A}V}(p^\prime)\rangle&=&
\lambda_{\widetilde{A}V}\,\varepsilon_\mu\, ,\nonumber \\
\langle 0| J_{-,\mu}^{\widetilde{V}A}(0)|Y_{\widetilde{V}A}(p^\prime)\rangle&=&
\lambda_{\widetilde{V}A}\,\varepsilon_\mu\, ,
\end{eqnarray}
\begin{eqnarray}\label{JSV-Residue}
\langle 0| J_{-,\mu\nu}^{S\widetilde{V}}(0)|Y_{S\widetilde{V}}(p^\prime)\rangle&=&
\lambda_{S\widetilde{V}}\,\varepsilon_{\mu\nu\alpha\beta}\,\varepsilon^\alpha p^{\prime\beta}\, , \nonumber\\
\langle 0| J_{-,\mu\nu}^{S\widetilde{V}}(0)|X_{S\widetilde{A}}(p^\prime)\rangle&=&
\bar{\lambda}_{S\widetilde{A}}\left(\varepsilon_{\mu}p^{\prime}_\nu -\varepsilon_{\nu}p^{\prime}_\mu\right)\, ,
\end{eqnarray}
the $\xi_\mu$ and $\varepsilon_\mu$ are the polarization vectors of the mesons and tetraquark states, respectively.

The explicit definitions for other hadronic coupling constants,
\begin{eqnarray}
\langle \bar{D}^*(p)D^*(q)|Y_{\widetilde{A}V}(p^\prime)\rangle&=&-\xi^*\cdot \xi^*(p-q)\cdot \varepsilon \,G_{\bar{D}^*D^*\widetilde{A}V}\, , \nonumber\\
\langle \bar{D}^*(p)D^*(q)|Y_{\widetilde{V}A}(p^\prime)\rangle&=&-\xi^*\cdot \xi^*(p-q)\cdot \varepsilon \,G_{\bar{D}^*D^*\widetilde{V}A}\, , \nonumber\\
\langle \bar{D}^*(p)D^*(q)|Y_{S\widetilde{V}}(p^\prime)\rangle&=&i\xi^*\cdot \xi^*(p-q)\cdot \varepsilon \,G_{\bar{D}^*D^*S\widetilde{V}}\, ,
\end{eqnarray}

\begin{eqnarray}
\langle \bar{D}_0(p)D^*(q)|Y_{\widetilde{A}V}(p^\prime)\rangle&=&\xi^*\cdot \varepsilon p\cdot q \,G_{\bar{D}_0D^*\widetilde{A}V}\, , \nonumber\\
\langle \bar{D}_0(p)D^*(q)|Y_{\widetilde{V}A}(p^\prime)\rangle&=&\xi^*\cdot \varepsilon  \,G_{\bar{D}_0D^*\widetilde{V}A}\, , \nonumber\\
\langle \bar{D}_0(p)D^*(q)|Y_{S\widetilde{V}}(p^\prime)\rangle&=&-i\xi^*\cdot \varepsilon  \,G_{\bar{D}_0D^*S\widetilde{V}}\, ,
\end{eqnarray}

\begin{eqnarray}
\langle \bar{D}_1(p)D(q)|Y_{\widetilde{A}V}(p^\prime)\rangle&=&i\xi^*\cdot \varepsilon \,G_{\bar{D}_1D\widetilde{A}V}\, , \nonumber \\
\langle \bar{D}_1(p)D(q)|Y_{\widetilde{V}A}(p^\prime)\rangle&=&i\xi^*\cdot \varepsilon p \cdot q\,G_{\bar{D}_1D\widetilde{V}A}\, , \nonumber \\
\langle \bar{D}_1(p)D(q)|Y_{S\widetilde{V}}(p^\prime)\rangle&=&\xi^*\cdot \varepsilon \,G_{\bar{D}_1DS\widetilde{V}}\, ,
\end{eqnarray}

\begin{eqnarray}
\langle \eta_c(p)\omega(q)|Y_{\widetilde{A}V}(p^\prime)\rangle&=&-\varepsilon^{\lambda\tau\rho\sigma}
q_\lambda \xi^*_\tau p^\prime_\rho \varepsilon_\sigma \,G_{\eta_c \omega \widetilde{A}V}\, , \nonumber \\
\langle \eta_c(p)\omega(q)|Y_{\widetilde{V}A}(p^\prime)\rangle&=&-\varepsilon^{\lambda\tau\rho\sigma}
q_\lambda \xi^*_\tau p^\prime_\rho \varepsilon_\sigma \,G_{\eta_c \omega \widetilde{V}A}\, ,\nonumber \\
\langle \eta_c(p)\omega(q)|Y_{S\widetilde{V}}(p^\prime)\rangle&=&-i\varepsilon^{\lambda\tau\rho\sigma}
q_\lambda \xi^*_\tau p^\prime_\rho \varepsilon_\sigma \,G_{\eta_c \omega S\widetilde{V}}\, , \nonumber \\
\langle \eta_c(p)\omega(q)|X_{S\widetilde{A}}(p^\prime)\rangle&=&-i \xi^* \cdot \varepsilon  \,\bar{G}_{\eta_c \omega S\widetilde{A}}\, ,
\end{eqnarray}

\begin{eqnarray}
\langle J/\psi(p)\omega(q)|Y_{\widetilde{A}V}(p^\prime)\rangle&=&\xi^*\cdot \xi^*(p-q)\cdot \varepsilon \,G_{J/\psi\omega\widetilde{A}V}\, , \nonumber \\
\langle J/\psi(p)\omega(q)|Y_{\widetilde{V}A}(p^\prime)\rangle&=&\xi^*\cdot \xi^*(p-q)\cdot \varepsilon \,G_{J/\psi\omega\widetilde{V}A}\, , \nonumber \\
\langle J/\psi(p)\omega(q)|Y_{S\widetilde{V}}(p^\prime)\rangle&=&i\xi^*\cdot \xi^*(p-q)\cdot \varepsilon \,G_{J/\psi\omega S\widetilde{V}}\, ,
\end{eqnarray}

\begin{eqnarray}
\langle \chi_{c0}(p)\omega(q)|Y_{\widetilde{A}V}(p^\prime)\rangle&=&-\xi^*\cdot \varepsilon \,G_{\chi_{c0}\omega\widetilde{A}V}\, , \nonumber \\
\langle \chi_{c0}(p)\omega(q)|Y_{\widetilde{V}A}(p^\prime)\rangle&=&\xi^*\cdot \varepsilon \,G_{\chi_{c0}\omega\widetilde{V}A}\, , \nonumber \\
\langle \chi_{c0}(p)\omega(q)|Y_{S\widetilde{V}}(p^\prime)\rangle&=&-i\xi^*\cdot \varepsilon \,G_{\chi_{c0}\omega S\widetilde{V}}\, ,
\end{eqnarray}

\begin{eqnarray}
\langle \chi_{c1}(p)\omega(q)|Y_{\widetilde{A}V}(p^\prime)\rangle&=&i
\varepsilon^{\rho\sigma\lambda\tau}p_\rho \xi_\sigma^* \xi^*_\lambda \varepsilon_\tau p \cdot q\,G_{\chi_{c1}\omega\widetilde{A}V}\, , \nonumber \\
\langle \chi_{c1}(p)\omega(q)|Y_{\widetilde{V}A}(p^\prime)\rangle&=&i
\varepsilon^{\rho\sigma\lambda\tau}p_\rho \xi_\sigma^* \xi^*_\lambda \varepsilon_\tau p \cdot q\,G_{\chi_{c1}\omega\widetilde{V}A}\, , \nonumber \\
\langle \chi_{c1}(p)\omega(q)|Y_{S\widetilde{V}}(p^\prime)\rangle&=&-
\varepsilon^{\rho\sigma\lambda\tau}p_\rho \xi_\sigma^* \xi^*_\lambda \varepsilon_\tau \,G_{\chi_{c1}\omega S\widetilde{V}}\, ,
\end{eqnarray}

\begin{eqnarray}
\langle J/\psi(p)f_0(q)|Y_{\widetilde{A}V}(p^\prime)\rangle&=&\xi^* \cdot \varepsilon  \,G_{J/\psi f_0\widetilde{A}V}\, , \nonumber \\
\langle J/\psi(p)f_0(q)|Y_{\widetilde{V}A}(p^\prime)\rangle&=&\xi^* \cdot \varepsilon  \,G_{J/\psi f_0\widetilde{V}A}\, , \nonumber \\
\langle J/\psi(p)f_0(q)|Y_{S\widetilde{V}}(p^\prime)\rangle&=&-i\xi^* \cdot \varepsilon  \,G_{J/\psi f_0S\widetilde{V}}\, .
\end{eqnarray}

The explicit expressions of the QCD side of other QCD sum rules,
\begin{eqnarray}
\Pi^{QCD}_{\bar{D}^*D\widetilde{A}V}(T^2)&=&\frac{m_c\langle\bar{q}g_s\sigma G q\rangle}{192\pi^2}\int_{m_c^2}^{s^0_D}du \left(1-\frac{3m_c^2}{u}\right)\frac{1}{u} \exp\left(-\frac{u+m_c^2}{T^2} \right)\nonumber\\
&&-\frac{m_c\langle\bar{q}g_s\sigma G q\rangle}{192\pi^2}\int_{m_c^2}^{s^0_{D^*}}ds \left(1+\frac{3m_c^2}{s}\right)\frac{1}{s} \exp\left(-\frac{s+m_c^2}{T^2} \right)\, ,
\end{eqnarray}

\begin{eqnarray}	 \Pi^{QCD}_{\bar{D}^*D^*\widetilde{A}V}(T^2)&=&\frac{m_c}{128\pi^4}
\int_{m_c^2}^{s^0_{D^*}}ds	\int_{m_c^2}^{s^0_{D^*}}du \left(1-\frac{m_c^2}{s}\right)^2\left(1-\frac{m_c^2}{u}\right)^2 \left(2u+m_c^2\right) \, \exp\left(-\frac{s+u}{T^2} \right) \nonumber\\
&&-\frac{m_c^2\langle\bar{q}q\rangle}{16\pi^2}\int_{m_c^2}^{s^0_{D^*}}ds \left(1-\frac{m_c^2}{s}\right)^2 \,\exp\left(-\frac{s+m_c^2}{T^2} \right)\nonumber\\
&&-\frac{\langle\bar{q}q\rangle}{48\pi^2}\int_{m_c^2}^{s^0_{D^*}}du \left(1-\frac{m_c^2}{u}\right)^2 \left(2u+m_c^2\right)\,\exp\left(-\frac{u+m_c^2}{T^2} \right)\nonumber\\
&&+\frac{\langle\bar{q}g_s\sigma G q\rangle}{576\pi^2T^2}\left( 4+\frac{3m_c^2}{T^2}\right)\int_{m_c^2}^{s^0_{D^*}}du\left( 1-\frac{m_c^2}{u}\right)^2 \left(2u+m_c^2\right)\, \exp\left(-\frac{u+m_c^2}{T^2} \right)\nonumber\\
&&+\frac{m_c^4\langle\bar{q}g_s\sigma G q\rangle}{64\pi^2T^4}\int_{m_c^2}^{s^0_{D^*}}ds \left(1-\frac{m_c^2}{s}\right)^2 \exp\left(-\frac{s+m_c^2}{T^2} \right)\nonumber\\
&&+\frac{m_c^2\langle\bar{q}g_s\sigma G q\rangle}{192\pi^2}\int_{m_c^2}^{s^0_{D^*}}ds \frac{1}{s}\left(1+\frac{4m_c^2}{s}\right)\, \exp\left(-\frac{s+m_c^2}{T^2} \right)\, ,
\end{eqnarray}

\begin{eqnarray}	 \Pi^{QCD}_{\bar{D}_0D^*\widetilde{A}V}(T^2)&=&\frac{3m_c^2}{128\pi^4}
\int_{m_c^2}^{s^0_{D_0}}ds 	\int_{m_c^2}^{s^0_{D^*}}du \left(1-\frac{m_c^2}{s}\right)^2\left(1-\frac{m_c^2}{u}\right)^2  \, \exp\left(-\frac{s+u}{T^2} \right) \nonumber\\
&&-\frac{m_c\langle\bar{q}q\rangle}{16\pi^2}\int_{m_c^2}^{s^0_{D_0}}ds \left(1-\frac{m_c^2}{s}\right)^2 \,\exp\left(-\frac{s+m_c^2}{T^2} \right)\nonumber\\
&&+\frac{m_c\langle\bar{q}q\rangle}{16\pi^2}\int_{m_c^2}^{s^0_{D^*}}du \left(1-\frac{m_c^2}{u}\right)^2 \,\exp\left(-\frac{u+m_c^2}{T^2} \right)\nonumber\\
&&-\frac{m_c^3\langle\bar{q}g_s\sigma G q\rangle}{64\pi^2T^4}\int_{m_c^2}^{s^0_{D^*}}du \left( 1-\frac{m_c^2}{u}\right)^2 \, \exp\left(-\frac{u+m_c^2}{T^2} \right)\nonumber\\
&&+\frac{m_c\langle\bar{q}g_s\sigma G q\rangle}{192\pi^2T^2}\left(4+\frac{3m_c^2}{T^2}\right)\int_{m_c^2}^{s^0_{D_0}}ds \left(1-\frac{m_c^2}{s}\right)^2 \exp\left(-\frac{s+m_c^2}{T^2} \right)\nonumber\\
&&-\frac{m_c\langle\bar{q}g_s\sigma G q\rangle}{192\pi^2}\int_{m_c^2}^{s^0_{D^*}}du \left(1+\frac{2m_c^2}{u}\right)\frac{1}{u}\, \exp\left(-\frac{u+m_c^2}{T^2} \right)	 \nonumber\\
&&+\frac{m_c\langle\bar{q}g_s\sigma G q\rangle}{192\pi^2}\int_{m_c^2}^{s^0_{D_0}}ds \left(1+\frac{2m_c^2}{s}\right)\frac{1}{s}\, \exp\left(-\frac{s+m_c^2}{T^2} \right)\, ,
\end{eqnarray}

\begin{eqnarray}	 \Pi^{QCD}_{\bar{D}_1D\widetilde{A}V}(T^2)&=&\frac{3}{128\pi^4}\int_{m_c^2}^{s^0_{D_1}}ds
\int_{m_c^2}^{s^0_D}du \left(1-\frac{m_c^2}{s}\right)^2\left(1-\frac{m_c^2}{u}\right)^2 \, u\, \left(2s+m_c^2\right) \, \exp\left(-\frac{s+u}{T^2} \right) \nonumber\\
&&-\frac{m_c\langle\bar{q}q\rangle}{16\pi^2}\int_{m_c^2}^{s^0_{D_1}}ds \left(1-\frac{m_c^2}{s}\right)^2\left(2s+m_c^2\right) \,\exp\left(-\frac{s+m_c^2}{T^2} \right)\nonumber\\
&&+\frac{3m_c\langle\bar{q}q\rangle}{16\pi^2}\int_{m_c^2}^{s^0_D}du \,u\,\left(1-\frac{m_c^2}{u}\right)^2 \,\exp\left(-\frac{u+m_c^2}{T^2} \right)\nonumber\\
&&-\frac{3m_c^3\langle\bar{q}g_s\sigma G q\rangle}{64\pi^2T^4}\int_{m_c^2}^{s^0_D}du\, u\,\left( 1-\frac{m_c^2}{u}\right)^2 \, \exp\left(-\frac{u+m_c^2}{T^2} \right)\nonumber\\
&&-\frac{m_c\langle\bar{q}g_s\sigma G q\rangle}{64\pi^2T^2}\left(2-\frac{m_c^2}{T^2}\right)\int_{m_c^2}^{s^0_{D_1}}ds \left(1-\frac{m_c^2}{s}\right)^2\left(2s+m_c^2\right)\exp\left(-\frac{s+m_c^2}{T^2} \right)\nonumber\\
&&-\frac{m_c\langle\bar{q}g_s\sigma G q\rangle}{384\pi^2}\int_{m_c^2}^{s^0_D}du \left(1-\frac{m_c^2}{u}\right)\left(3-\frac{m_c^2}{u}\right)\, \exp\left(-\frac{u+m_c^2}{T^2} \right)	\nonumber\\
&&+\frac{m_c^3\langle\bar{q}g_s\sigma G q\rangle}{192\pi^2}\int_{m_c^2}^{s^0_{D_1}}ds \frac{1}{s}\, \exp\left(-\frac{s+m_c^2}{T^2} \right)\nonumber\\
&&-\frac{m_c\langle\bar{q}g_s\sigma G q\rangle}{384\pi^2}\int_{m_c^2}^{s^0_D}du \left(3-\frac{m_c^4}{u^2}\right)\, \exp\left(-\frac{u+m_c^2}{T^2} \right)\, ,
\end{eqnarray}

\begin{eqnarray}
\Pi^{QCD}_{\eta_c\omega\widetilde{A}V}(T^2)
&=&\frac{m_c\langle\bar{q}q\rangle}{2\sqrt{2}\pi^2}\int_{4m_c^2}^{s^0_{\eta_c}}ds
\frac{\sqrt{\lambda(s,m_c^2,m_c^2)}}{s}  \, \exp\left(-\frac{s}{T^2} \right)\nonumber\\
&&-\frac{m_c\langle\bar{q}g_s\sigma G q\rangle}{6\sqrt{2}\pi^2T^2}\int_{4m_c^2}^{s^0_{\eta_c}}ds
\frac{\sqrt{\lambda(s,m_c^2,m_c^2)}}{s}  \, \exp\left(-\frac{s}{T^2} \right)\nonumber\\
&&-\frac{m_c\langle\bar{q}g_s\sigma G q\rangle}{8\sqrt{2}\pi^2}\int_{4m_c^2}^{s^0_{\eta_c}}ds
\frac{1}{\sqrt{s(s-4m_c^2)}}  \, \exp\left(-\frac{s}{T^2} \right)\, ,
\end{eqnarray}

\begin{eqnarray}
\Pi^{QCD}_{J/\psi\omega\widetilde{A}V}(T^2)&=&0\, ,
\end{eqnarray}

\begin{eqnarray}
\Pi^{QCD}_{\chi_{c0}\omega\widetilde{A}V}(T^2)&=&\frac{3}{32\sqrt{2}\pi^4}
\int_{4m_c^2}^{s^0_{\chi_{c0}}}ds
\int_{0}^{s^0_\omega}du \frac{\sqrt{\lambda(s,m_c^2,m_c^2)}}{s}\, u \left(s-4m_c^2\right)  \, \exp\left(-\frac{s+u}{T^2} \right)\, ,
\end{eqnarray}

\begin{eqnarray}
\Pi^{QCD}_{\chi_{c1}\omega\widetilde{A}V}(T^2)&=&
\frac{\langle\bar{q}q\rangle}{12\sqrt{2}\pi^2}\int_{4m_c^2}^{s^0_{\chi_{c1}}}ds
 \frac{\sqrt{\lambda(s,m_c^2,m_c^2)}}{s}\frac{s+2m_c^2}{s}  \exp\left(-\frac{s}{T^2} \right)\nonumber\\
&&-\frac{\langle\bar{q}g_s\sigma G q\rangle}{36\sqrt{2}\pi^2T^2}\int_{4m_c^2}^{s^0_{\chi_{c1}}}ds
\frac{\sqrt{\lambda(s,m_c^2,m_c^2)}}{s}\frac{s+2m_c^2}{s}  \exp\left(-\frac{s}{T^2} \right)\, ,	
\end{eqnarray}

\begin{eqnarray}
\Pi^{QCD}_{J/\psi f_0\widetilde{A}V}(T^2)&=&\frac{3}{32\sqrt{2}\pi^4}\int_{4m_c^2}^{s^0_{J/\psi}}ds
\int_{0}^{s^0_{f_0}}du \frac{\sqrt{\lambda(s,m_c^2,m_c^2)}}{s}\,u\left(s+2m_c^2\right)  \exp\left(-\frac{s+u}{T^2} \right)\, ,
\end{eqnarray}

\begin{eqnarray}
\Pi^{QCD}_{\bar{D}D\widetilde{V}A}(T^2)&=&\frac{\langle\bar{q}g_s\sigma G q\rangle}{384\pi^2}\int_{m_c^2}^{s^0_D}du \left(1-\frac{m_c^2}{u}\right)\left(3+\frac{m_c^2}{u}\right)\, \exp\left(-\frac{u+m_c^2}{T^2} \right)\nonumber\\
&&-\frac{m_c^2\langle\bar{q}g_s\sigma G q\rangle}{64\pi^2}\int_{m_c^2}^{s^0_D}ds \left(1-\frac{m_c^2}{s}\right)\frac{1}{s} \exp\left(-\frac{s+m_c^2}{T^2} \right)\nonumber\\
&&+\frac{\langle\bar{q}g_s\sigma G q\rangle}{384\pi^2}\int_{m_c^2}^{s^0_D}du \left(3+\frac{m_c^4}{u^2}\right) \exp\left(-\frac{u+m_c^2}{T^2} \right)\, ,
\end{eqnarray}

\begin{eqnarray}
 \Pi^{QCD}_{\bar{D}^*D\widetilde{V}A}(T^2)&=&\frac{3m_c^2}{128\pi^4}
 \int_{m_c^2}^{s^0_{D^*}}ds\int_{m_c^2}^{s^0_D}du \left(1-\frac{m_c^2}{s}\right)^2\left(1-\frac{m_c^2}{u}\right)^2  \, \exp\left(-\frac{s+u}{T^2} \right) \nonumber\\
&&-\frac{m_c\langle\bar{q}q\rangle}{16\pi^2}\int_{m_c^2}^{s^0_{D^*}}ds \left(1-\frac{m_c^2}{s}\right)^2 \,\exp\left(-\frac{s+m_c^2}{T^2} \right)\nonumber\\
&&-\frac{m_c\langle\bar{q}q\rangle}{16\pi^2}\int_{m_c^2}^{s^0_D}du \left(1-\frac{m_c^2}{u}\right)^2 \,\exp\left(-\frac{u+m_c^2}{T^2} \right)\nonumber\\
&&+\frac{m_c\langle\bar{q}g_s\sigma G q\rangle}{192\pi^2T^2}\left( 4+\frac{3m_c^2}{T^2}\right)\int_{m_c^2}^{s^0_D}du \left( 1-\frac{m_c^2}{u}\right)^2  \exp\left(-\frac{u+m_c^2}{T^2} \right)\nonumber\\
&&+\frac{m_c^3\langle\bar{q}g_s\sigma G q\rangle}{64\pi^2T^4}\int_{m_c^2}^{s^0_{D^*}}ds \left(1-\frac{m_c^2}{s}\right)^2 \exp\left(-\frac{s+m_c^2}{T^2} \right)\nonumber\\
&&+\frac{m_c\langle\bar{q}g_s\sigma G q\rangle}{192\pi^2}\int_{m_c^2}^{s^0_D}du \left(1+\frac{2m_c^2}{u}\right)\frac{1}{u}\, \exp\left(-\frac{u+m_c^2}{T^2} \right)	 \nonumber\\
&&+\frac{m_c\langle\bar{q}g_s\sigma G q\rangle}{192\pi^2}\int_{m_c^2}^{s^0_{D^*}}ds \frac{1}{s}\, \exp\left(-\frac{s+m_c^2}{T^2} \right) \, ,
\end{eqnarray}

\begin{eqnarray}
\Pi^{QCD}_{\bar{D}^*D^*\widetilde{V}A}(T^2)&=&-\frac{\langle\bar{q}g_s\sigma G q\rangle}{384\pi^2}\int_{m_c^2}^{s^0_{D^*}}du \left(1-\frac{m_c^2}{u}\right) \left(3+\frac{m_c^2}{u}\right)\,\exp\left(-\frac{u+m_c^2}{T^2} \right)\nonumber\\
&&+\frac{m_c^2\langle\bar{q}g_s\sigma G q\rangle}{192\pi^2}\int_{m_c^2}^{s^0_{D^*}}ds \left(1-\frac{m_c^2}{s}\right)\frac{1}{s} \exp\left(-\frac{s+m_c^2}{T^2} \right)\nonumber\\
&&+\frac{\langle\bar{q}g_s\sigma G q\rangle}{384\pi^2}\int_{m_c^2}^{s^0_{D^*}}du \left(3+\frac{5m_c^4}{u^2}\right)\, \exp\left(-\frac{u+m_c^2}{T^2} \right)\, ,
\end{eqnarray}

\begin{eqnarray}
 \Pi^{QCD}_{\bar{D}_0D^*\widetilde{V}A}(T^2)&=&\frac{3}{128\pi^4}
 \int_{m_c^2}^{s^0_{D_0}}ds	\int_{m_c^2}^{s^0_{D^*}}du \left(1-\frac{m_c^2}{s}\right)^2\left(1-\frac{m_c^2}{u}\right)^2\, s\, \left(2u+m_c^2\right)  \exp\left(-\frac{s+u}{T^2} \right) \nonumber\\
&&-\frac{3m_c\langle\bar{q}q\rangle}{16\pi^2}\int_{m_c^2}^{s^0_{D_0}}ds \,s\,\left(1-\frac{m_c^2}{s}\right)^2 \,\exp\left(-\frac{s+m_c^2}{T^2} \right)\nonumber\\
&&+\frac{m_c\langle\bar{q}q\rangle}{16\pi^2}\int_{m_c^2}^{s^0_{D^*}}du \left(2u+m_c^2\right)\left(1-\frac{m_c^2}{u}\right)^2\exp\left(-\frac{u+m_c^2}{T^2} \right)\nonumber\\
&&+\frac{m_c\langle\bar{q}g_s\sigma G q\rangle}{64\pi^2T^2}\left(2-\frac{m_c^2}{T^2} \right)\int_{m_c^2}^{s^0_{D^*}}du \left( 1-\frac{m_c^2}{u}\right)^2\left(2u+m_c^2\right) \exp\left(-\frac{u+m_c^2}{T^2} \right)\nonumber\\
&&+\frac{3m_c^3\langle\bar{q}g_s\sigma G q\rangle}{64\pi^2T^4}\int_{m_c^2}^{s^0_{D_0}}ds \,s\left(1-\frac{m_c^2}{s}\right)^2 \exp\left(-\frac{s+m_c^2}{T^2} \right)\nonumber\\
&&+\frac{m_c\langle\bar{q}g_s\sigma G q\rangle}{192\pi^2}\int_{m_c^2}^{s^0_{D_0}}ds \left(3-\frac{2m_c^2}{s}\right) \exp\left(-\frac{s+m_c^2}{T^2} \right)\nonumber\\
&&-\frac{m_c^3\langle\bar{q}g_s\sigma G q\rangle}{192\pi^2}\int_{m_c^2}^{s^0_{D^*}}du\, \frac{1}{u}\, \exp\left(-\frac{u+m_c^2}{T^2} \right) \, ,
\end{eqnarray}

\begin{eqnarray}
 \Pi^{QCD}_{\bar{D}_1D\widetilde{V}A}(T^2)&=&\frac{3m_c^2}{128\pi^4}
 \int_{m_c^2}^{s^0_{D_1}}ds	\int_{m_c^2}^{s^0_D}du \left(1-\frac{m_c^2}{s}\right)^2\left(1-\frac{m_c^2}{u}\right)^2  \, \exp\left(-\frac{s+u}{T^2} \right) \nonumber\\
&&-\frac{m_c\langle\bar{q}q\rangle}{16\pi^2}\int_{m_c^2}^{s^0_{D_1}}ds \left(1-\frac{m_c^2}{s}\right)^2 \,\exp\left(-\frac{s+m_c^2}{T^2} \right)\nonumber\\
&&+\frac{m_c\langle\bar{q}q\rangle}{16\pi^2}\int_{m_c^2}^{s^0_D}du \left(1-\frac{m_c^2}{u}\right)^2 \,\exp\left(-\frac{u+m_c^2}{T^2} \right)\nonumber\\
&&-\frac{m_c\langle\bar{q}g_s\sigma G q\rangle}{192\pi^2T^2}\left( 4+\frac{3m_c^2}{T^2}\right)\int_{m_c^2}^{s^0_D}du \left( 1-\frac{m_c^2}{u}\right)^2 \, \exp\left(-\frac{u+m_c^2}{T^2} \right)\nonumber\\
&&+\frac{m_c^3\langle\bar{q}g_s\sigma G q\rangle}{64\pi^2T^4}\int_{m_c^2}^{s^0_{D_1}}ds \left(1-\frac{m_c^2}{s}\right)^2 \exp\left(-\frac{s+m_c^2}{T^2} \right)\nonumber\\
&&-\frac{m_c\langle\bar{q}g_s\sigma G q\rangle}{192\pi^2}\int_{m_c^2}^{s^0_D}du \left(1+\frac{2m_c^2}{u}\right)\frac{1}{u}\, \exp\left(-\frac{u+m_c^2}{T^2} \right)	 \nonumber\\
&&+\frac{m_c\langle\bar{q}g_s\sigma G q\rangle}{192\pi^2}\int_{m_c^2}^{s^0_{D_1}}ds \left(1+\frac{2m_c^2}{s}\right)\frac{1}{s}\, \exp\left(-\frac{s+m_c^2}{T^2} \right)\, ,
\end{eqnarray}

\begin{eqnarray}
 \Pi^{QCD}_{\eta_c\omega\widetilde{V}A}(T^2)&=&
 \frac{m_c\langle\bar{q}q\rangle}{2\sqrt{2}\pi^2}
 \int_{4m_c^2}^{s^0_{\eta_c}}ds\frac{\sqrt{\lambda(s,m_c^2,m_c^2)}}{s}   \exp\left(-\frac{s}{T^2} \right)\nonumber\\
&&-\frac{m_c\langle\bar{q}g_s\sigma G q\rangle}{6\sqrt{2}\pi^2T^2}\int_{4m_c^2}^{s^0_{\eta_c}}ds\frac{\sqrt{\lambda(s,m_c^2,m_c^2)}}{s}
\exp\left(-\frac{s}{T^2} \right)\nonumber\\
&&-\frac{m_c\langle\bar{q}g_s\sigma G q\rangle}{8\sqrt{2}\pi^2}\int_{4m_c^2}^{s^0_{\eta_c}}ds
\frac{1}{\sqrt{s(s-4m_c^2)}}   \exp\left(-\frac{s}{T^2} \right)\, ,
\end{eqnarray}

\begin{eqnarray}
\Pi^{QCD}_{J/\psi\omega\widetilde{V}A}(T^2)&=&0\, ,
\end{eqnarray}

\begin{eqnarray}
 \Pi^{QCD}_{\chi_{c0}\omega\widetilde{V}A}(T^2)&=&\frac{3}{32\sqrt{2}\pi^4}
 \int_{4m_c^2}^{s^0_{\chi_{c0}}}ds	\int_{0}^{s^0_\omega}du \frac{\sqrt{\lambda(s,m_c^2,m_c^2)}}{s}\,u\left(s-4m_c^2\right)  \exp\left(-\frac{s+u}{T^2} \right)\, ,
\end{eqnarray}

\begin{eqnarray}
 \Pi^{QCD}_{\chi_{c1}\omega\widetilde{V}A}(T^2)&=&
 \frac{\langle\bar{q}q\rangle}{12\sqrt{2}\pi^2}\int_{4m_c^2}^{s^0_{\chi_{c1}}}ds
\frac{\sqrt{\lambda(s,m_c^2,m_c^2)}}{s}\frac{s+2m_c^2}{s} \exp\left(-\frac{s}{T^2} \right)\nonumber\\
&&-\frac{\langle\bar{q}g_s\sigma G q\rangle}{36\sqrt{2}\pi^2T^2}\int_{4m_c^2}^{s^0_{\chi_{c1}}}ds
\frac{\sqrt{\lambda(s,m_c^2,m_c^2)}}{s}\frac{s+2m_c^2}{s}  \exp\left(-\frac{s}{T^2} \right)\, ,
\end{eqnarray}

\begin{eqnarray}
\Pi^{QCD}_{J/\psi f_0\widetilde{V}A}(T^2)&=&\frac{3}{32\sqrt{2}\pi^4}\int_{4m_c^2}^{s^0_{J/\psi}}
ds	\int_{0}^{s^0_{f_0}}du \frac{\sqrt{\lambda(s,m_c^2,m_c^2)}}{s}\,u\left(s+2m_c^2\right)  \exp\left(-\frac{s+u}{T^2} \right)\, ,
\end{eqnarray}

\begin{eqnarray}
\Pi^{QCD}_{\bar{D}DS\widetilde{V}}(T^2)&=&\frac{3m_c^2}{256\pi^4}
\int_{m_c^2}^{s^0_D}ds	\int_{m_c^2}^{s^0_D}du \left(1-\frac{m_c^2}{s}\right)^2\left(1-\frac{m_c^2}{u}\right)^2  \exp\left(-\frac{s+u}{T^2} \right) \nonumber\\
&&-\frac{m_c\langle\bar{q}q\rangle}{32\pi^2}\int_{m_c^2}^{s^0_D}ds \left(1-\frac{m_c^2}{s}\right)^2 \exp\left(-\frac{s+m_c^2}{T^2} \right)\nonumber\\
&&-\frac{m_c\langle\bar{q}q\rangle}{32\pi^2}\int_{m_c^2}^{s^0_D}du \left(1-\frac{m_c^2}{u}\right)^2 \exp\left(-\frac{u+m_c^2}{T^2} \right)\nonumber\\
&&+\frac{m_c^3\langle\bar{q}g_s\sigma G  q\rangle}{128\pi^2T^4}\int_{m_c^2}^{s^0_D}ds \left(1-\frac{m_c^2}{s}\right)^2 \exp\left(-\frac{s+m_c^2}{T^2} \right)\nonumber\\
&&+\frac{m_c^3\langle\bar{q}g_s\sigma G q\rangle}{128\pi^2T^4}\int_{m_c^2}^{s^0_D}du \left(1-\frac{m_c^2}{u}\right)^2\, \exp\left(-\frac{u+m_c^2}{T^2} \right)\nonumber\\
&&-\frac{m_c\langle\bar{q}g_s\sigma G q\rangle}{384\pi^2}\int_{m_c^2}^{s^0_D}ds \left(2-\frac{m_c^2}{s}\right)\frac{1}{s} \exp\left(-\frac{s+m_c^2}{T^2} \right)\nonumber\\
&&-\frac{m_c\langle\bar{q}g_s\sigma G q\rangle}{384\pi^2}\int_{m_c^2}^{s^0_D}du \left(2-\frac{m_c^2}{u}\right)\frac{1}{u} \exp\left(-\frac{u+m_c^2}{T^2} \right)\, ,
\end{eqnarray}

\begin{eqnarray}
\Pi^{QCD}_{\bar{D}^*DS\widetilde{V}}(T^2)&=&
\frac{m_c^4\langle\bar{q}g_s\sigma G q\rangle}{96\pi^2}\int_{m_c^2}^{s^0_{D^*}}ds \frac{1}{s^3} \exp\left(-\frac{s+m_c^2}{T^2} \right)\, ,
\end{eqnarray}

\begin{eqnarray}
\Pi^{QCD}_{\bar{D}^*D^* S\widetilde{V}}(T^2)&=&-\frac{m_c^3\langle\bar{q}g_s\sigma G q\rangle}{128\pi^2}\int_{m_c^2}^{s^0_{D^*}}du \frac{1}{u^2} \exp\left(-\frac{u+m_c^2}{T^2} \right)\nonumber\\
&&-\frac{m_c^3\langle\bar{q}g_s\sigma G q\rangle}{192\pi^2}\int_{m_c^2}^{s^0_{D^*}}ds \frac{1}{s^2} \exp\left(-\frac{s+m_c^2}{T^2} \right) \, ,
\end{eqnarray}

\begin{eqnarray}
 \Pi^{QCD}_{\bar{D}_0D^*S\widetilde{V}}(T^2)&=&\frac{m_c}{128\pi^4}
 \int_{m_c^2}^{s^0_{D_0}}ds	\int_{m_c^2}^{s^0_{D^*}}du \left(1-\frac{m_c^2}{s}\right)^2\left(1-\frac{m_c^2}{u}\right)^2 \left(2u+m_c^2\right)  \exp\left(-\frac{s+u}{T^2} \right) \nonumber\\
&&-\frac{m_c^2\langle\bar{q}q\rangle}{16\pi^2}\int_{m_c^2}^{s^0_{D_0}}ds \left(1-\frac{m_c^2}{s}\right)^2 \exp\left(-\frac{s+m_c^2}{T^2} \right)\nonumber\\
&&+\frac{\langle\bar{q}q\rangle}{48\pi^2}\int_{m_c^2}^{s^0_{D^*}}du \left(1-\frac{m_c^2}{u}\right)^2 \left(2u+m_c^2\right) \,\exp\left(-\frac{u+m_c^2}{T^2} \right)\nonumber\\
&&-\frac{m_c^2\langle\bar{q}g_s\sigma G q\rangle}{192\pi^2T^4}\int_{m_c^2}^{s^0_{D^*}}du \left( 1-\frac{m_c^2}{u}\right)^2\left(2u+m_c^2\right) \exp\left(-\frac{u+m_c^2}{T^2} \right)\nonumber\\
&&+\frac{m_c^4\langle\bar{q}g_s\sigma G q\rangle}{64\pi^2T^4}\int_{m_c^2}^{s^0_{D_0}}ds \left(1-\frac{m_c^2}{s}\right)^2 \exp\left(-\frac{s+m_c^2}{T^2} \right)\nonumber\\
&&-\frac{m_c^2\langle\bar{q}g_s\sigma G q\rangle}{192\pi^2}\int_{m_c^2}^{s^0_{D_0}}ds \left(2-\frac{m_c^2}{s}\right)\frac{1}{s} \exp\left(-\frac{s+m_c^2}{T^2} \right)\nonumber\\
&&+\frac{m_c^4\langle\bar{q}g_s\sigma G q\rangle}{192\pi^2}\int_{m_c^2}^{s^0_{D^*}}du \frac{1}{u^2} \exp\left(-\frac{u+m_c^2}{T^2} \right)\, ,
\end{eqnarray}

\begin{eqnarray}
 \Pi^{QCD}_{\bar{D}_1DS\widetilde{V}}(T^2)&=&\frac{m_c}{128\pi^4}\int_{m_c^2}^{s^0_{D_1}}ds
\int_{m_c^2}^{s^0_D}du \left(1-\frac{m_c^2}{s}\right)^2\left(1-\frac{m_c^2}{u}\right)^2 \left(2s+m_c^2\right)  \exp\left(-\frac{s+u}{T^2} \right) \nonumber\\
&&-\frac{\langle\bar{q}q\rangle}{48\pi^2}\int_{m_c^2}^{s^0_{D_1}}ds \left(1-\frac{m_c^2}{s}\right)^2 \left(2s+m_c^2\right) \exp\left(-\frac{s+m_c^2}{T^2} \right)\nonumber\\
&&+\frac{m_c^2\langle\bar{q}q\rangle}{16\pi^2}\int_{m_c^2}^{s^0_D}du \left(1-\frac{m_c^2}{u}\right)^2  \exp\left(-\frac{u+m_c^2}{T^2} \right)\nonumber\\
&&-\frac{m_c^4\langle\bar{q}g_s\sigma G q\rangle}{64\pi^2T^4}\int_{m_c^2}^{s^0_D}du \left( 1-\frac{m_c^2}{u}\right)^2   \exp\left(-\frac{u+m_c^2}{T^2} \right)\nonumber\\
&&+\frac{m_c^2\langle\bar{q}g_s\sigma G q\rangle}{192\pi^2T^4}\int_{m_c^2}^{s^0_{D_1}}ds \left(1-\frac{m_c^2}{s}\right)^2 \left(2s+m_c^2\right) \exp\left(-\frac{s+m_c^2}{T^2} \right)\nonumber\\
&&-\frac{m_c^4\langle\bar{q}g_s\sigma G q\rangle}{192\pi^2}\int_{m_c^2}^{s^0_{D_1}}ds \frac{1}{s^2}  \exp\left(-\frac{s+m_c^2}{T^2} \right)\nonumber\\
&&+\frac{m_c^2\langle\bar{q}g_s\sigma G q\rangle}{192\pi^2}\int_{m_c^2}^{s^0_D}du \left(2-\frac{m_c^2}{u}\right)\frac{1}{u}  \exp\left(-\frac{u+m_c^2}{T^2} \right)\, ,
\end{eqnarray}

\begin{eqnarray}
\Pi^{QCD}_{\eta_c\omega S\widetilde{V}}(T^2)&=&\frac{m_c}{16\sqrt{2}\pi^4}\int_{4m_c^2}^{s^0_{\eta_c}}ds
\int_{0}^{s^0_\omega}du \,u\,\frac{\sqrt{\lambda(s,m_c^2,m_c^2)}}{s} \exp\left(-\frac{s+u}{T^2} \right)\, ,
\end{eqnarray}

\begin{eqnarray}
\Pi^{QCD}_{J/\psi\omega S\widetilde{V}}(T^2)&=&\frac{m_c\langle\bar{q}q\rangle}{4\sqrt{2}\pi^2}
\int_{4m_c^2}^{s^0_{J/\psi}}ds
\frac{\sqrt{\lambda(s,m_c^2,m_c^2)}}{s} \exp\left(-\frac{s}{T^2} \right)\nonumber\\
&&-\frac{m_c\langle\bar{q}g_s\sigma G q\rangle}{12\sqrt{2}\pi^2T^2}\int_{4m_c^2}^{s^0_{J/\psi}}ds
\frac{\sqrt{\lambda(s,m_c^2,m_c^2)}}{s} \exp\left(-\frac{s}{T^2} \right)\nonumber\\
&&+\frac{m_c\langle\bar{q}g_s\sigma G q\rangle}{48\sqrt{2}\pi^2}\int_{4m_c^2}^{s^0_{J/\psi}}ds
\frac{2s-m_c^2}{s\sqrt{s(s-4m_c^2)}}  \exp\left(-\frac{s}{T^2} \right)\, ,
\end{eqnarray}

\begin{eqnarray}
\Pi^{QCD}_{\chi_{c0}\omega S\widetilde{V}}(T^2)&=&\frac{\langle\bar{q}q\rangle}{4\sqrt{2}\pi^2}
 \int_{4m_c^2}^{s^0_{\chi_{c0}}}ds
\frac{\sqrt{\lambda(s,m_c^2,m_c^2)}}{s}\left(s-4m_c^2\right)  \exp\left(-\frac{s}{T^2} \right)\nonumber\\
&&-\frac{\langle\bar{q}g_s\sigma G q\rangle}{12\sqrt{2}\pi^2T^2}\int_{4m_c^2}^{s^0_{\chi_{c0}}}ds
\frac{\sqrt{\lambda(s,m_c^2,m_c^2)}}{s}\left(s-4m_c^2 \right)   \exp\left(-\frac{s}{T^2} \right)\nonumber\\
&&+\frac{\langle\bar{q}g_s\sigma G q\rangle}{24\sqrt{2}\pi^2}\int_{4m_c^2}^{s^0_{\chi_{c0}}}ds
\frac{s-5m_c^2}{\sqrt{s(s-4m_c^2)}}  \exp\left(-\frac{s}{T^2} \right)\, ,
\end{eqnarray}

\begin{eqnarray}
\Pi^{QCD}_{\chi_{c1}\omega S\widetilde{V}}(T^2)&=&\frac{1}{48\sqrt{2}\pi^4}\int_{4m_c^2}^{s^0_{\chi_{c1}}}ds
\int_{0}^{s^0_\omega}du \frac{\sqrt{\lambda(s,m_c^2,m_c^2)}}{s}\,u\left(s-4m_c^2\right)   \exp\left(-\frac{s+u}{T^2} \right)\, ,
\end{eqnarray}

\begin{eqnarray}
\Pi^{QCD}_{J/\psi f_0 S\widetilde{V}}(T^2)&=&\frac{3m_c}{32\sqrt{2}\pi^4}\int_{4m_c^2}^{s^0_{J/\psi}}ds
	\int_{0}^{s^0_{f_0}}du \,u\frac{\sqrt{\lambda(s,m_c^2,m_c^2)}}{s} \exp\left(-\frac{s+u}{T^2} \right)\, ,
\end{eqnarray}
$\lambda(a,b,c)=a^2+b^2+c^2-2ab-2bc-2ca$.

\section*{Acknowledgements}
This  work is supported by National Natural Science Foundation, Grant Number  12175068.

\end{document}